\documentclass{ws-ijmpd2}
\usepackage[super,compress]{cite}

\usepackage{moresize}
\usepackage{enumerate}
\usepackage[caption=false]{subfig}
\usepackage{multirow}
\usepackage{mathtools}
\usepackage{stackengine}

\newcommand{\be}{\begin{equation}}
\newcommand{\ee}{\end{equation}}
\newcommand{\bea}{\begin{eqnarray}}
\newcommand{\eea}{\end{eqnarray}}
\newcommand{\bse}{\begin{subequations}}
\newcommand{\ese}{\end{subequations}}
\newcommand{\bce}{\begin{center}}
\newcommand{\ece}{\end{center}}
\newcommand{\bfg}{\begin{figure}}
\newcommand{\efg}{\end{figure}}
\newcommand{\bed}{\begin{description}}
\newcommand{\eed}{\end{description}}
\newcommand{\ben}{\begin{enumerate}}
\newcommand{\een}{\end{enumerate}}

\newcommand{\bit}{\begin{itemlist}}
\newcommand{\eit}{\end{itemlist}}

\newcommand{\nn}{\nonumber}

\newcommand{\pa}{\partial}
\newcommand{\fr}{\frac}
\newcommand{\sq}{\sqrt}
\newcommand{\no}{\noindent}

\def\a  {\alpha}
\def\b  {\beta}
\def\c  {\gamma}

\def\d  {\delta}

\def\f  {\phi}
\def\F  {\Phi}
\def\k  {\kappa}

\def\L  {\Lambda}
\def\m  {\mu}
\def\n  {\nu}

\def\O  {\Omega}

\def\r  {\rho}
\def\th {\theta}
\def\s  {\sigma}

\def\vph {\varphi}

\def\le {\left}
\def\ri {\right}
\newcommand{\cA}{\mathcal A}

\newcommand{\cH}{\mathcal H}
\newcommand{\cL}{\mathcal L}

\newcommand{\cQ}{\mathcal Q}
\newcommand{\cR}{\mathcal R}
\newcommand{\cS}{\mathcal S}

\newcommand{\fw}{\mathfrak w}
\newcommand{\nab}{\nabla}

\newcommand{\hl}{\widehat{\lambda}}
\newcommand{\hg}{\widehat{g}}

\newcommand{\hT}{\widehat{T}}

\newcommand{\hR}{\widehat{\cR}}
\newcommand{\hS}{\widehat{\cS}}
\newcommand{\hLm}{\widehat{\cL}^{(m)}}

\newcommand{\hnab}{\widehat{\nabla}}

\newcommand{\dmt}{\d^{(m)}}
\newcommand{\dmp}{\d^{(m)}_{_0}}
\newcommand{\tmt}{\th^{(m)}}
\newcommand{\rmt}{\r^{(m)}}

\newcommand{\rft}{\r^{(\vph)}}
\newcommand{\pft}{p^{(\vph)}}

\newcommand{\Om}{\O^{(m)}}

\newcommand{\rmp}{\r^{(m)}_{_0}}
\newcommand{\Omp}{\O^{(m)}_{_0}}

\newcommand{\sw}{\mathsf w}

\newcommand{\Hp}{H_{_0}}
\newcommand{\fp}{\f_{_0}}
\newcommand{\tp}{t_{_0}}

\newcommand{\se}{\s^{(8)}}
\newcommand{\sep}{\s^{(8)}_{_0}}

\newcommand*\rfra[2]{{}^{\scriptstyle{#1}}\!\!\diagup_{\!\!\scriptstyle{#2}}}

\newcommand{\bdm}{\begin{displaymath}}
\newcommand{\edm}{\end{displaymath}}

\long\def\symbolfootnote[#1]#2{\begingroup%
\def\thefootnote{\fnsymbol{footnote}}\footnote[#1]{#2}\endgroup}
\numberwithin{equation}{section}
\providecommand{\appendixname}{appendix}

\begin{document}

\markboth{Mohit Kumar Sharma and Sourav Sur}
{Imprints of interacting dark energy on...}

%
%

\title{Imprints of interacting dark energy on cosmological perturbations }

\author{MOHIT KUMAR SHARMA}

\address{\it Department of Physics \& Astrophysics, University of Delhi\\
New Delhi - 110 007, India\\
mr.mohit254@gmail.com}

\author{SOURAV SUR}

\address{\it Department of Physics \& Astrophysics, University of Delhi\\
New Delhi - 110 007, India\\
sourav.sur@gmail.com; sourav@physics.du.ac.in}

\maketitle

\begin{history}
\received{}
\revised{}
\end{history}

\begin{abstract}
We investigate the characteristic modifications in the evolving 
cosmological perturbations when dark energy interacts with dust-like 
matter, causing the latter's background energy density fall off with 
time faster than usual. Focusing in particular to the late-time 
cosmic evolution, we show that such an interaction (of a specific 
form, arising naturally in a scalar-tensor formulation, or a wide 
range of modified gravity equivalents thereof), can have a rather 
significant effect on the perturbative spectrum, than on the background 
configuration which is not expected to get distorted much from $\L$CDM. 
Specifically, the matter density contrast, which is by and large 
scale-invariant in the deep sub-horizon limit, not only gets dragged 
as the interaction affects the background Hubble expansion rate, but 
also receives a contribution from the perturbation in the (scalar 
field induced) dark energy, which oscillates about a non-zero mean 
value. As such, the standard parametrization ansatz for the the matter 
density growth factor becomes inadequate. So we modify it suitably, 
and also find a numerical fit of the growth index in terms of the 
background parameters, in order to alleviate the problems that arise 
otherwise. Such a fit enables direct estimations of the background 
parameters, as well as the growth parameter and the reduced Hubble 
parameter, which we duly carry out using a redshift space distortion 
(RSD) subsample and its combination with the observational Hubble data.
On the whole, the parametric estimates show consistency with the 
general observational constraints on the background level cosmology,
as well as the constraints on scalar-tensor gravity from astrophysical 
observations, apart from having significance in the domain of 
cosmological perturbations. 
\end{abstract}


\section{Introduction}  \label{sec:intro}

One of the main compulsions in modern cosmology is to determine the 
extent to which the rate of formation of large-scale structures (LSS) 
is affected by the late-time cosmic acceleration, and the constraints 
imposed thereby on the dynamics of the dark energy (DE) which 
supposedly drives this acceleration
\cite{CST-rev,FTH-rev,AT-book,wols-ed,MCGM-ed}.
As the observations strongly favour the DE component to be a 
cosmological constant $\L$, there is an obvious dilution of any idea 
circumventing the dynamical evolution of the same. Nevertheless, such 
a dynamics may be reckoned, at least, within the proximity of {\it mild} 
($1\s$ level) distortions in the parametric estimations for the 
concordant $\L$CDM model, with $\L$ and cold dark matter (CDM) as the 
dominant components of the universe
\cite{wmap9-CP,wmap9-Fin,bet-SN,Planck15-CP,Planck15-DEMG,
scol-SN,Planck18-CP}.
There lies a caveat though, which is even more suggestive in the 
sense that the effect of something ostensibly mild at the background 
cosmological level, and at low redshifts ($z \lesssim 1$), may not 
be so at a perturbative level and at moderate to high redshifts. It 
therefore necessitates one to test independently the mildness of the 
DE dynamics, sought in a given cosmological model, from perturbative 
studies extending to large enough redshifts. Admittedly, the analysis 
of the observational data on the growth of LSS (or that of the matter 
density perturbations) can place robust constraints on a given DE 
model, and possibly discriminate it among various other models.

Commonly known dynamical models of DE (quintessence, kessence, etc.
\cite{CRS-quin,CLW-quin,tsuj-quin,AMS2000-kess,AMS2001-kess,
MCLT-kess,schr-kess,SSSD-kess,SS-dquin}),
inspite of their merits, are plugged with some serious issues which 
make it imperative to search for wayouts in alternative scenarios. 
An area of considerable interest, in this respect, is that of the 
{\em modified gravity} (MG) theories
\cite{NO-mgDE,tsuj-mgDE,CFPS-mg,JBE-mgDE,papa-ed,NOO-mg},
which go beyond the realm of General Relativity (GR)
\cite{FM-ST,frni-ST,BP-ST,BEPS-ST,TUMTY-ST,ENO-ST,CHL-ST,BGP-ST,
ST-ST,ST-KT}. 
Such theories not only provide a geometric interpretation of the DE 
in the standard cosmological setup, but also alleviate the {\em cosmic 
coincidence} problem in the equivalent scalar-tensor formulations, by 
emulating effective scenarios in which the DE component can interact 
with matter fields (the CDM sources inclusive)
%
%
\cite{CST-rev,AT-book,WS-intDE,amend-intDE,CPR-intDE,FP-intDE,CW-intDE,
CHOP-intDE,CHP2008-intDE,CHP2009-intDE,PSC-intDE,GGQ-intDE,BKLY-intDE,
land-intDE,TM-intDE,amend-rev,ban-intDE}.
In fact, many phenomenological studies involving the DE-matter (DEM) 
interactions draw proper physical explanation from scalar-tensor 
equivalent MG models, including those which lead to the coveted
revelation of a unified cosmic dark sector
\cite{TM-intDE,amend-rev,BBM-uDE,BBPP-uDE,GNP-uDE,FFKB-uDE,
CMV-mm,MV-mm,NOO-mfR,SVM-mm,LMNV-mdhost,CM-mmg1,CM-mmg2,CMR-maf,CDS-MMT,
SDC-MMT}.

Scalar-tensor formulations provide a natural perception of the DEM 
interactions, by virtue of the effective contact coupling(s) of the 
scalar field source(s) of the DE and matter field(s) in the conformally 
transformed frames
\cite{FM-ST,frni-ST,BP-ST,BEPS-ST,TUMTY-ST,ENO-ST,CHL-ST,BGP-ST,ST-ST}.
A given such interaction affects the evolution of the total background 
matter density $\rmt(z)$, as well as the growth of the corresponding 
perturbation $\d\rmt(z)$, and hence the growth of the LSS
\cite{koiv-grow,PA-grow,PG-grow,GP-grow,amen-pert,SHCK-grow}. 
More specifically, the interaction (with a scalar field $\vph$) makes 
$\rmt(z)$ drift from its usual ({\em dust}-like) form\footnote{For 
simplicity, in this paper, we consider matter content of the universe
to be very nearly pressure-free. In particular, apart from the CDM, 
we consider only the visible matter in the form of a dust of baryons,
the pressure due to which is negligibly small compared to the corresponding 
energy density and is safely ignorable even at the linear perturbative 
cosmological level.}, thereby leading to a drag force on $\d\rmt(z)$. 
Consequently, the evolution profile of the matter density contrast 
$\,\dmt(z) := \d\rmt(z)/\rmt(z) \,$ and the corresponding growth factor 
$f(z)$ differ from those in the non-interacting scenarios, such as 
quintessence. The field perturbation $\d\vph$, on the other hand, 
undergoes a damped oscillatory evolution in the sub-horizon regime, 
quite similar to that in the non-interacting scenarios. However, the 
oscillations are about a non-zero mean value, proportional to the 
strength of the $\vph$-coupling with matter. As such, $\d\vph$ 
contributes to the matter velocity divergence, affecting in turn the 
evolution of $\dmt(z)$ by a further extent
\cite{AT-book,koiv-grow}.

From the technical point of view, an intriguing outcome of a DEM 
interaction is that the matter perturbation growth factor $f(z)$ can 
acquire a value $> 1$ at large $z$, contrary to the restriction set
in the non-interacting scenarios. So there is a stringent need to 
look beyond the commonly known $f(z)$ parametrizations which prohibit 
the breach of the $f(z) = 1$ barrier at any phase of evolution of the 
universe
\cite{PA-grow,PG-grow,GP-grow,WYF-grwpara,PAB-grwpara,BBS-grwpara}.
In particular, it is worth looking for a suitable modification of the 
parametrization $f(z) = \le[\Om(z)\ri]^{\c(z)}$, where $\Om(z)$ is the 
matter density parameter and $\c(z)$ is the so-called {\em growth index}, 
which is not a constant even for $\L$CDM
\cite{PG-grow,GP-grow,WYF-grwpara,PAB-grwpara,BBS-grwpara}.
This parametrization is well-motivated and widely used in the literature,
since the assertion of the form of the function $\c(z)$, and its value 
$\c_{_0}$ at the present epoch ($z = 0$), from the Red-shift space 
distortion (RSD) observations, proves to be a convenient way of comparing 
cosmological models of various sort, as well as checking the viability of 
the same.
\cite{PG-grow,GP-grow,WYF-grwpara,PAB-grwpara,BBS-grwpara,BP-grwindx,
SBM-grwindx,bat-grwindx,MBMDR-grwindx,PSG-grwindx,BA-grwindx,HKV-modsel,
SS-MSTintDE}. 
%
%
%

In attempting the modification of such a parametrization, in presence 
of a DEM interaction, our objective in this paper is to determine the 
observational constraints on the background level parameters, such as 
$\Omp = \Om\big\vert_{z=0}$ and the DEM coupling strength $n$, in 
a scalar-tensor cosmological scenario. Nevertheless, we keep our 
attention to a scalar-tensor configuration of a specific sort, which 
is essentially meant for a typical case study in this paper, with a 
common motivation from the perspective of the equivalence with a wide 
range of MG formulations in the literature. Besides, we focus on 
studying the evolution of $f(z)$ only at the sub-horizon scales, at 
which the late-time growth of the LSS is relevant observationally. 
However, our endeavor is to perform the growth analysis in a general
way, without undermining the role of the field perturbation $\d\vph$, 
as is quite often done in practice. Afterall, as mentioned above, the 
oscillations of $\d\vph$ are about a non-zero mean value in presence 
of a DEM interaction. So, depending on the strength of the latter, 
there can be an effect of some significance on $f(z)$, and hence on 
the growth index $\c(z)$, especially in the deep sub-horizon regime. 

Now, the $\c(z)$ stipulations in the literature are mostly power series 
expansions, about $z = 0$, $\,\Om = 1$, and so on
\cite{PG-grow,GP-grow,WYF-grwpara,PAB-grwpara,BBS-grwpara,BP-grwindx,
SBM-grwindx}.
Although effective, they have limited scope of applicability, for 
instance, only at very late times, or only in the deep matter-dominated
era. Moreover, the power series coefficients increase the parameter 
space of the model, unless their dependence on the background level 
model parameters is explicitly worked out. It is nonetheless desirable 
to solve the evolution equation for $f(z)$ directly by using the 
latter's parametrization, in terms of the growth index $\c(z)$. Doing 
this analytically is however a hard proposition. So, one may look for 
a numerical fit of $\c(z)$ in terms of the background parameters, and 
check the consistency upto large redshifts. We attempt such a numerical 
fitting, in order to carry out subsequently the direct estimations of 
the parameters using the RSD observational data (or the GOLD sub-sample 
thereof
\cite{SNS-GR}), 
and also the latter's combination with the observational Hubble ($H(z)$) 
data
\cite{RDR-bao}. 
Apart from $n$ and $\Omp$, there are two parameters involved, viz. 
reduced Hubble constant $h$ and the RSD parameter $\, \sep \equiv 
\se\big\vert_{z=0}\,$, where $\se(z)$ is the root-mean-square 
fluctuation of the mass distribution within a sphere of radius 
$8\,$Mph$^{-1}$.

The paper is organized as follows: in section \ref{sec:DEM-formalism}, 
we first recapitulate how the cosmological scenarios emerging from
scalar-tensor theories (or their MG equivalents) naturally accommodate 
the DEM interactions, and then demonstrate the background cosmological 
solution, given in a parametric form, for a specific class of such 
theories (detailed in the Appendix). We go on to study the corresponding
cosmological perturbations thereafter, in section \ref{sec:cosm-pert}, 
and obtain the evolution equations for $\dmt(z)$, $f(z)$ and $\d\vph(z)$, 
in the well-known Newtonian gauge. Numerically solving those equations, 
for certain fiducial parametric settings, we examine the extent to 
which the effect of the DEM interaction on the evolution of $f(z)$ is 
accountable, as opposed to the latter's (negligible) scale-dependence, 
at the sub-horizon scales. We follow this up, in section \ref{sec:Growth}, 
with a proposed parametrization ansatz for $f(z)$ in presence of the DEM
interaction, and subsequently obtain the requisite numerical fit of the 
growth index $\c(z)$ in terms of the background parameters. The statistical 
estimations of the requisite parameters $\le(n, \Omp, \sep, h\ri)$ are 
carried out next, in section \ref{sec:Param-est}, using the 
Metropolis-Hastings algorithm for the Markov Chain Monte Carlo (MCMC) 
simulation with the two chosen datasets mentioned above. Finally, we 
summarize our findings and conclude in section \ref{sec:concl}.

\bigskip

\no 
{\large \sl Conventions and Notations}: Throughout this paper, we use metric 
signature $\, (-,+,+,+)$ and natural units, with the speed of light $c = 1$.
We denote and the gravitational coupling factor by $\, \k = \sq{8 \pi G_N}$, 
where $G_N$ is the Newton's constant, the metric determinant by $g$, and the 
values of parameters or functions at the present epoch by an affixed subscript 
`$0$'.

\section{Interacting Dark Energy-Matter scenario} \label{sec:DEM-formalism}

Let us recall that in GR the Einstein tensor $G_{\m\n}$ gets restricted by 
the (contracted) Bianchi identity $\nab^\m \, G_{\m\n} = 0$, which implies 
the consistency of the conservation relation for the energy-momentum tensor 
$T_{\m\n}$, i.e.
\be \label{T-consv}
\nab^\m\, T_{\m\n} =\, 0 \, ,
\ee
with the Einstein's equations $G_{\m\n} = \k^2 \,T_{\m\n}\,$.
However, a flexibility is there in the GR formulation itself. That is, for a 
multi-component system (such as the universe consisting of radiation, baryonic 
matter, CDM, DE etc.), the conservation of $T_{\m\n}$ does not necessarily imply 
the same for each of the component energy-momentum tensors $T^{(i)}_{\m\n}$. 
Therefore, re-expressing Eq.\,(\ref{T-consv}) in the form
\be \label{T_comp}
\nab^\m \, T^{(i)}_{\m\n} =\, Q^{(i)}_\n \quad \mbox{with} \quad 
\sum_i Q^{(i)}_\n = 0 \, ,
\ee
one can always ponder on the scenarios in which some, or all, of these components 
may have mutual interactions, quantified in terms of the vectors $Q^{(i)}_\n$.

Now, from the cosmological perspective, it is reasonable to consider the 
evolution of the universe (particularly at late times, i.e. at moderate 
to low redshifts) to be driven by two interacting components, viz. a 
pressure-less matter component (comprized of the visible baryons and the 
CDM) and a DE component (presumably induced by some scalar field $\vph$). 
Eq.\,(\ref{T_comp}) then implies that the respective energy-momentum 
tensors, $T^{(m)}_{\m\n}$ and $T^{(\vph)}_{\m\n}$, satisfy the conservation 
relations  
\be \label{T_m_DE}
\nab^\m \, T^{(m)}_{\m\n} =\, Q_\n \quad \mbox{and} \quad 
\nab^\m \, T^{(\vph)}_{\m\n} =\, - \, Q_\n \,,
\ee
with the vector $Q_\n$ determining the extent of the 
interaction\footnote{There could be interactions within the 
matter sector as well, i.e. among the baryons and various dark 
matter species. However, for simplicity, we exclude such a 
possibility in this work.}. Such an interaction is by no means 
ad-hoc --- it appears naturally in scalar-tensor theories, 
which are characterized by explicit non-minimal gravitational 
couplings with effective scalar degree(s) of freedom in the 
Jordan frame. As is well-known, a suitable conformal 
transformation of the Jordan frame metric can lift the 
non-minimality, however, at the expense of leaving the 
effective matter Lagrangian dependent on the corresponding 
scalar field $\vph$, both implicitly and explicitly, in the 
Einstein frame. While making interpretations in the standard 
Friedmann-Robertson-Walker (FRW) cosmological setup, such a 
$\vph$-dependent matter Lagrangian can at once be identified 
as that which invokes an energy exchange between the 
($\vph$-induced) DE and the matter sector, or in other 
words, a DEM interaction. 

In principle, depending on the type of non-minimal coupling of the 
gravitational Lagrangian $R$ (or the Ricci curvature scalar) and 
a scalar field ($\f$, say) in the Jordan frame, there can be a 
variety of functional forms of the interaction vector $Q_\m$ 
appearing in the Einstein frame. For definiteness however, in 
this paper we shall consider a scalar-tensor formulation of a 
particular sort, marked by a $\f^2 R$ term in the Jordan frame 
action, which then corresponds to the Brans-Dicke action (with 
a possible augmentation of a potential term for the scalar field). 
After a conformal transformation and a field re-definition 
$\f = \k^{-1} e^{n \k \vph}$, one finds  
\be \label{Int-vec}
\cQ_\m =\, \k \, n \, \rmt \pa _\m \vph \,,
\ee
where $\rmt$ is the matter density defined in the Einstein 
frame\footnote{The physical entities are assumed to be definable 
and existent in the Einstein frame, and the cosmological matter 
under consideration is the pressure-less (baryonic + dark) matter, 
with the trace of the corresponding energy-momentum tensor equal 
to $\,- \rmt$.}, and $\,n = (6 + \fw)^{-1/2}$, with $\fw$ denoting 
the Brans-Dicke parameter (see the Appendix for details). 

This sort of formulation is equivalent to that of a wide range of 
modified or alternative theories of gravity, albeit with a fixed 
value of the parameter $n$ in some cases. For instance, the $f(R)$
theories are well-known to have Brans-Dicke equivalent formulations
only for $\fw = 0$, i.e. $\,n = \rfra 1 {\!\sq{6}}\,$ (fixed). We 
shall treat such fixed $n$ theories as exceptions though, or in 
other words, consider $n$ as a free parameter throughout the rest 
of this paper. In fact, we shall estimate the value of $n$ (or 
determine its upper bound) using observational data, in the 
cosmological perturbative analysis due to be carried out in the 
subsequent sections.

Refer now to the standard spatially flat FRW space-time
geometry, described by the scale factor $a(t)$ and the Hubble 
parameter $H(t) = \dot a(t)/a(t) \,$, where the overdot $\{\cdot\}$ 
denotes differentiation with respect to the cosmic time $t$. In 
accord with our presumption of the universe being comprized of a
pressure-free matter component and a $\vph$-induced DE component, 
we have the Friedmann equation and the conservation relations (that 
follow from Eqs.\,(\ref{T_m_DE}) and (\ref{Int-vec})) given by 
\bea 
&& H^2 =\, \fr{\k^2} 3 \le[\rmt +\, \rft\ri] \,, \label{FRW-eq} \\
&& {\dot\r}^{(m)} +\, 3 H \,\r^{(m)}  
=\, -\, \k \, n \, \rmt \, \dot\vph \,\,, \label{FRW-consv_m} \\
&& \dot{\r}^{(\vph)} +\, 3 H \le[\rft +\, \pft\ri] 
=\, \k \, n \, \rmt \, \dot\vph \,\,, \label{FRW-consv_f}
\eea
with the usual expressions for the energy density and pressure
\be \label{DE-dens-pr}
\rft =\, \fr{{\dot \vph}^2} 2 +\, U(\vph) \quad \mbox{and} \quad
\pft =\, \fr{{\dot \vph}^2} 2 -\, U(\vph) \,\,.
\ee
corresponding to the field $\vph$, with a potential $U(\vph)$.

Eq.\,(\ref{FRW-consv_m}) at once gives 
\be \label{rho_m}
\rmt(a) \propto\, a^{-3} e^{- \k n \vph(a)} \,\,,
\ee
which qualitatively shows how the matter density deviates from 
its usual dust-like form ($\sim a^{-3}$) because of the 
interaction. However, the determination of the complete 
solutions for $\rmt(a)$, $\rft(a)$ and $\pft(a)$ requires one 
to have the prior knowledge of the functional form of the 
potential $U(\vph)$. 

Commonly known forms of $U(\vph)$ in the literature, for which 
exact analytic cosmological solutions have been worked out, are 
certain single or double exponentials 
\cite{GGQ-intDE,BKLY-intDE,land-intDE,TM-intDE}.
However, many such exponential forms lack proper physical 
motivation, and many such solutions involve quite a few free 
parameters, neither of which are desirable from theoretical 
perspectives. So, for a concrete case study in this paper, we 
shall limit ourselves to an exact solution of a specific sort, 
given by the set
\cite{SSASB-MST}: 
\bea 
&& \vph(a) =\, 2 n \, \k^{-1} \ln a \,, \label{f-sol} \\
&& \rmt(a) =\, \rmp a^{-(3+2n^2)} \,, \label{rho_m-sol} \\
&& \r^{(\vph)}(a) =\, 2 n^2 \k^{-2} H^2 (a) 
+\, \L \, a^{-4n^2} \,, \label{rho_f-sol} \\
&& p^{(\vph)}(a) =\, 2 n^2 \k^{-2} H^2 (a) 
-\, \L \, a^{-4n^2} \,, \label{p_f-sol}
\eea
where $\L$ is a constant of mass dimension $= 4$, $\,\rmp$ is
the value of the matter density at the present epoch ($t = \tp$, 
whence $a = 1$), and
\be \label{Hub-sol}
H(a) =\, \fr \k {\sq{3-2n^2}} \, \le[\fr{\rmp}{a^{3+2n^2}} 
+\, \fr \L {a^{4n^2}}\ri]^{1/2} \,.
\ee
This solution set is useful from the following points of view:
\bit
\item It involves no more than two free parameters, since 
one parameter in the set $\le[n, \L, \rmp\ri]$ always gets 
determined by the other two, by virtue of the constraint 
$\O_{_0} = 1$, where $\O_{_0}$ is the total density parameter 
of the universe at the present epoch $\tp$. 
\item It reduces to the standard $\L$CDM solution just as the
coupling parameter $n \rightarrow 0$, or equivalently, $\fw 
\rightarrow \infty$. 
Therefore, considering the solar system constraints on the 
scalar-tensor equivalent theories of gravity, which essentially 
imply a large lower bound on $\fw$, we may infer that the 
dynamical evolution of the DE, due to a non-vanishing $n$ here, 
has to be slow enough.
\item It can be obtained under the assertion of a single 
exponential potential
%
\be \label{Pot}
U(\vph) =\, \L\, e^{- 2 \k n \vph} \,\,,
\ee
which amounts to just a mass term for the scalar field $\f$ in 
the Jordan frame. The mass parameter $m = 2 \k^2 \L$, which one 
may verify easily (from the general formalism detailed in the 
Appendix). As a typical exemplary scenario, we may refer to that 
of the metric-scalar-torsion theory, in which such a mass term 
effectively originates from a norm-fixing condition on the axial 
mode vector $\cA^\m$ of torsion, or from a higher order term 
$\le(\cA_\m \cA^\m\ri)^2$ augmented to the Lagrangian
\cite{SSASB-MST,ASBSS-MSTda,ASBSS-MSTpp}. 
\eit
From the above Eqs.\,(\ref{rho_m-sol})\,--\,(\ref{Hub-sol}), one
derives the matter density parameter and the total equation of 
state (EoS) parameter of the system, respectively, as
\bea 
&& \Om(a) :=\, \fr{\rmt}{\rft + \rmt} =\, 
\fr{\le(3 - 2n^2\ri)\Omp}{\le(3 - 2n^2 - 3\Omp\ri) 
a^{3 - 2n^2} +\, 3\Omp} \,\,, \label{Om} \\
&& \sw(a) =\, \fr{\pft}{\rft + \rmt} =\, \Om(a) \,-\, 1 
+\, \fr{4n^2} 3  \,, \label{EoS}
\eea
where $\Omp$ is the value of $\Om$ at the present epoch.


\section{Evolution of cosmological perturbations} \label{sec:cosm-pert}

Let us study, in this section, the evolution of cosmological 
perturbations in the well-known Newtonian gauge in which the 
perturbed line element is given by
\be \label{pert-ds}
ds^2 =\, e^{2N} \Big[- \le(1 - 2\F\ri) \cH^{-2} dN^2 +\, 
\le(1 + 2\F\ri) \d_{ij} dx^i dx^j\Big] \,,
\ee
where $N(t) = \ln a(t)$ is the number of $e$-foldings, $\cH(t) 
= a(t) H(t)$ is the conformal Hubble parameter and $\F$ is the 
Bardeen potential\footnote{Due to the absence of anisotropic stress 
at late-times, we are free to consider both the temporal and spatial 
perturbations to be described by $\F$
\cite{CG-LSS,BLPS-DSA}.
}. For convenience, we shall use the units $\k^2 = 8 \pi G_N = 1$, 
and denote all the derivatives by subscripted commas (e.g. $\,d\F/dN 
\equiv \F_{,_N} \,, ~ d^2 \F/dN^2 \equiv \F_{,_{NN}}\,,~ dU/d\vph 
\equiv U_{,\vph}\,$, etc.), from here on. 

Given a fluid, which is assumed to be {\it perfect} even at a 
perturbative level, we have the perturbation in the corresponding 
energy-momentum tensor expressed as  
\be \label{pert-T}
\d T^\m_{~\n} = \big[(1 + c_s^2)\, u^\m u_\n +\, c_s^2\, 
\d^\m_\n\big] \d\r \,+ \le(p + \r\ri) \big(u^\m \d u_\n +\, 
u^\n \d u_\m\big) \,, 
\ee
where $u^\a, \r$ and $p$ denote the background fluid velocity, 
energy density and pressure, with the respective perturbations 
$\d u^\a, \d\r$ and $\d p$, and $c_s^2 := dp/d\r$ defines the 
squared sound speed of the perturbations.

Resort now to the system of two interacting components, viz. 
the universe, with the DE induced by a scalar field $\vph$, 
and the matter sector considered as a pressure-less fluid, 
presumably, at all levels of perturbation, which implies that 
the corresponding sound speed $c_s = 0$. Recall the equation 
(\ref{FRW-consv_m}) which shows how the conservation of the 
background matter density $\rmt$ is affected by the coupling 
with $\vph$. Perturbing this equation, and following the 
standard procedure of dealing with the Fourier transforms 
of all the pertubations involved, 
we derive the conservation equation for the matter density 
perturbation $\d \rmt$ as
\be \label{dmt}
\dmt_{,_N} =\, - \le[\tmt +\, 3 \F_{,_N}\ri] - \,
n\, (\d\vph)_{,_N} \,\,,
\ee
where $\dmt = \d \rmt/\rmt$ is the matter density contrast,
and $\d\vph$ and $\tmt$ respectively denote the scalar field 
perturbation and the matter velocity divergence
\be \label{tmt}
\tmt_{,_N} =\, -\, \fr{\tmt} 2 \Big(1 -\, 3 \sw -\, 2n \,
\vph_{,_N}\Big) -\, \fr 1 {\hl^2} \Big(\F +\, n \d\vph\Big)\,,
\ee
where $\sw$ is the background EoS parameter given by 
Eq.\,(\ref{EoS}), and $\, \hl \equiv \cH/ k$, with $k$ 
being the comoving wavenumber\footnote{To be specific, 
Eq.\,(\ref{dmt}) follows from the conservation of $\d T^0_{~0}$ 
component of Eq.\,(\ref{T_m_DE}), whereas the Euler equation 
(\ref{tmt}) is obtained from that of the $\d T^0_{~i}$ component,
in addition to the use of the form the perturbed metric 
($\ref{pert-ds}$) and the background matter conservation equation
(\ref{FRW-consv_m}). Of course, the final forms of the expressions
are derived by expanding each perturbed quantity, say $h(\vec{x},N)$, 
in the Fourier space, viz. $h(\vec{x},N) = \int d^3k \, h(\vec{k},N) 
e^{i \vec{k}\cdot \vec{x}}$.}.

From Eqs.\,(\ref{dmt}) and (\ref{tmt}) one obtains the second 
order differential equation
\bea \label{dmt-k}
\dmt_{,_{NN}} &=& -\, \fr 1 2 \Big[3\F_{,_N} +\, \dmt_{,_N}
+\, n (\d\vph)_{,_N}\Big] \Big(1 -\, 3 \sw -\, 2n \,
\vph_{,_N}\Big) \nn\\
&& \qquad \qquad
-\, 3 \F_{,_{NN}} -\, n (\d\vph)_{,_{NN}} +\, \fr 1 {\hl^2} 
\Big(\F +\, n \d\vph\Big)\,,
\eea
which shows that the dependence of $\dmt$ on the coupling 
parameter $n$ is compounded with its scale ($\hl$) dependence,
because of that of the Bardeen potential $\F$. However, more 
strikingly, both these dependencies have an added influence of 
$\d\vph$ which depends on $n$ and $\hl$ as well. This can be
inferred from the equation  
\bea \label{dphi-k}
&& (\d\vph)_{,_{NN}} \,+ \le(2 + \fr{\cH_{,_{N}}}{\cH}\ri) 
(\d\vph)_{,_{N}} + \le[\fr 1 {\hl^2} +\, \fr{(U_{,\vph\vph})^2}
{H^2}\ri] \d\vph \nn \\ 
&& \qquad =\, 3n \le(\dmt - 2\F\ri) \Om \,+\, \fr{2 \F \,U_{,\vph}}
{H^2} \,+\, 4 \F_{,_{N}} \vph_{,_{N}} \,\,,
\eea
which one obtains in a similar way (by perturbing the field 
conservation equation (\ref{FRW-consv_f}), and taking note that 
$c_s^2 = 1$ for the field perturbations)\footnote{The sound speed 
of field perturbations refers to the coefficient of the second 
order spatial derivative of $\vph$. In the Fourier space, the 
coefficient of $\hl^{-2} \d\vph$ determines that sound speed, which
identically turns out to be unity for the perturbed configuration
we are dealing here.}.

On the other hand, the scale-dependence of $\F$ is evident from 
the time-time and time-space components of the perturbed Einstein's 
equations, viz.
\bea  
&& \F_{,_{N}} +\, \F =\, - \fr 1 2 \bigg[\vph_{,_{N}} \d\vph \,+\, 
3 \hl^2 \,\Om \tmt\bigg]\,, \label{dF-k} \\
&& \bigg(1 +\, \fr{\hl^2 U}{H^2}\bigg) \F \,=\, \fr{\hl^2} 2 
\bigg[3 \Om \dmt \,+\, \fr{U_{,\vph}}{H^2} \d\vph \nn\\
&& \qquad \qquad \qquad \qquad \qquad
+ \Big\{(\d\vph)_{,_{N}} +\, 3 \d\vph\Big\} \vph_{,_{N}} +\,
9 \hl^2 \, \Om \tmt\bigg] \,. \label{F-k}
\eea
In fact, together with Eq.\,(\ref{dmt-k}), these equations make it 
clear that $\F$ is scale-dependent, and so is $\dmt$, even when 
there is no scalar field $\vph$. It also follows that in such a 
case the scale-dependence of $\dmt$ can be ignored only in the 
deep sub-horizon regime ($k \gg \cH\,$ or $\,\hl \ll 1$), whence 
$\F$ becomes negligible. This is of course well-known, and can be 
inferred in presence of the field $\vph$ as well. However, the 
scenario is rather intriguing when $\vph$ couples to matter 
non-minimally, and thereby leads to the DEM interaction. One may 
then ask the natural question as to {\em whether such a coupling 
can be of any significance at all, in a study of cosmological 
perturbations limited to the sub-horizon scales only}. That is 
to say, {\em whether the dependence of $\dmt$ on the coupling 
strength (or the parameter $n$) can lead to a stronger effect 
on its evolution profile, than that due to the scale-dependence, 
which is nonetheless suppressed in the deep sub-horizon regime}.

We endeavor to address this here, by first using Eqs.\,(\ref{dF-k}) 
and (\ref{F-k}) to eliminate $\F$ and its derivatives from 
Eqs.\,(\ref{dmt-k}) and (\ref{dphi-k}), and then solving the 
resulting equations numerically, for certain chosen values of $n$ 
and $\hl$, at least an order of magnitude below unity. As the 
background cosmological solution, we consider the one given by the 
set (\ref{f-sol})\,-\,(\ref{Hub-sol}) and (\ref{Om}), (\ref{EoS}), 
corresponding to the chosen potential (\ref{Pot}). Resorting then to 
the fiducial parametric settings 
$$
\Omp = 0.3\,, \quad h \equiv H_0/[100 \,\mbox{Km s}^{-1} 
\mbox{Mpc}^{-1}] = 0.67 \,,
$$
and the adiabatic initial conditions, viz. 
$$
\dmt \simeq \dmt_{_{,N}} \simeq e^N \simeq 0.0009 \quad 
\mbox{and} \quad \d\vph \simeq \d\vph_{_{,N}} \simeq 0
$$
at the recombination epoch $N \simeq - 7$ (i.e. at redshift 
$\,z \equiv e^{-N} - 1 \simeq 1100$), we obtain the evolution 
profiles of $\dmt$ and $\d\vph$ for those choices of $n$ and 
$\hl$. 

\begin{figure}[htp!]
\centering
\subfloat[{\footnotesize $\dmt$ and $\d\vph$ for a fixed $n$ and 
different $\hl$}]
{\includegraphics[scale=0.6]{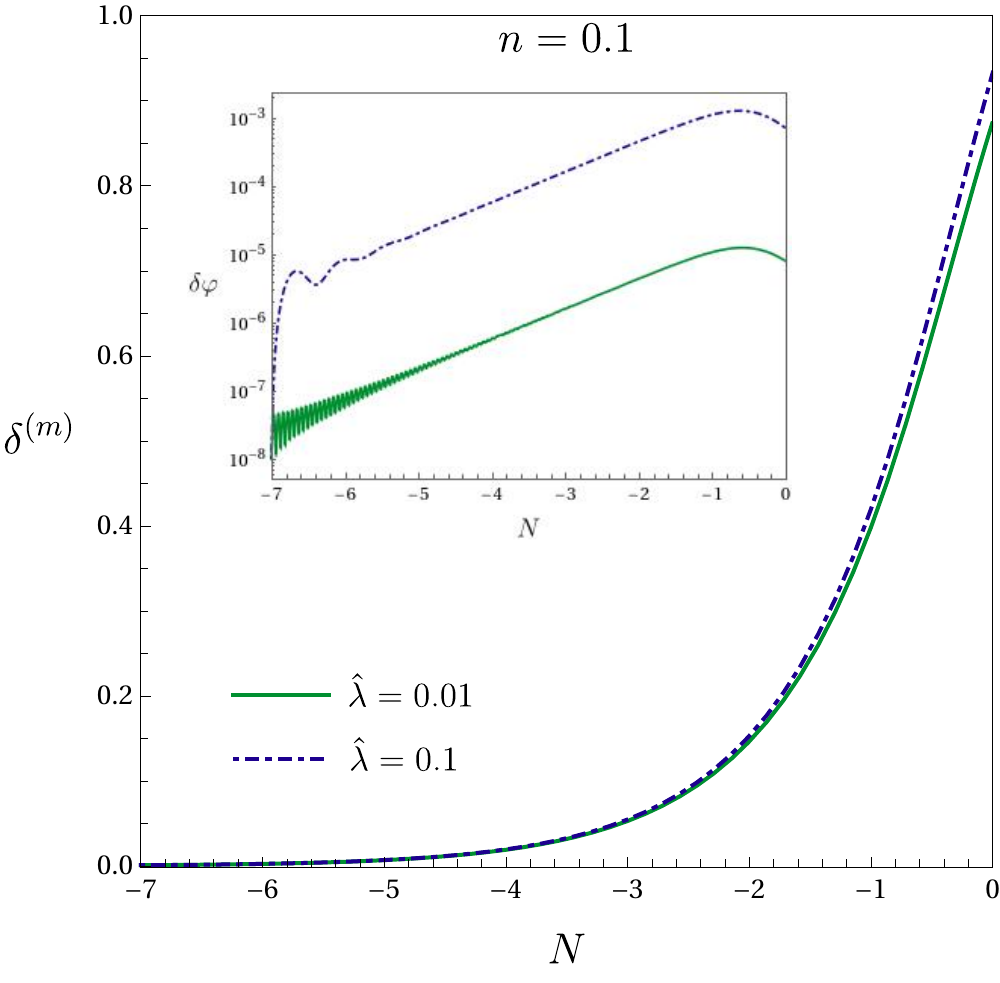}} \quad
\subfloat[{\footnotesize $\dmt$ and $\d\vph$ for a fixed $\hl$ and 
different $n$}]
{\includegraphics[scale=0.6]{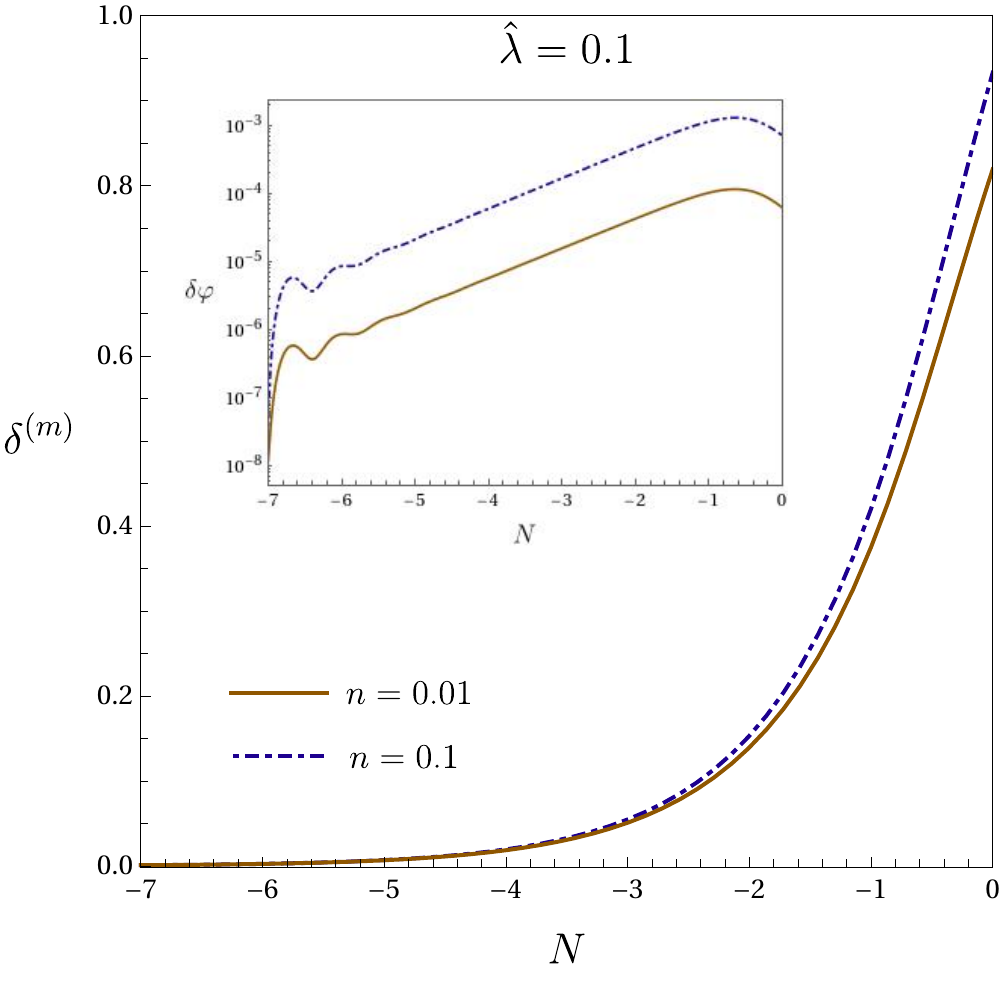}}  
\caption{\footnotesize Evolution profiles of the matter density 
contrast $\dmt(N)$ in the range $N \in [-7,0]$, for (a) fixed 
coupling strength $n = 0.1$ and different perturbation scales 
$\hl = 0.01, 0.1$, and (b) fixed perturbation scale $\hl = 0.1$ 
and different coupling strengths $n = 0.01, 0.1$. The insets 
show the respective (log scale) evolution profiles of the field 
perturbation $\d\vph(N)$, in the same range.}
\label{fig:dmdf_l_n}
\end{figure}

Figs.\,\ref{fig:dmdf_l_n}\,(a) and (b) depict the evolution of
$\dmt$ in the range $N \in [-7,0]$, respectively, for $n = 0.1$ 
(fixed) and two values of $\hl \,(= 0.01, 0.1)$, and for $\hl 
= 0.1$ (fixed) and two values of $n \,(= 0.01, 0.1)$.
%
%
The evolution profile is exponential, which gets steeper with the 
increase of $\hl$ from $0.01$ to $0.1$, or with the increase of $n$ 
by the same amount, in the respective cases. However, the margin of 
such an enhancement in the steepness is quite less in the former, 
than in the latter, as noticed particularly from the deviation at 
late times (i.e. close to the present epoch $N=0$). This at least 
shows qualitatively that a very nearly scale-independent evolution 
of $\dmt$ at small scales does not necessarily imply that the effect 
on it due to a non-minimal $\vph$-matter coupling is ignorable.  

The insets of Figs.\,\ref{fig:dmdf_l_n}\,(a) and (b) show the 
corresponding evolution profiles of $\d\vph$, characterized by 
initial oscillations, with a gradual smoothening out, due to 
damping, as $N$ increases. The initial oscillations however become 
less rapid as $\hl$ is increased from $0.01$ to $0.1$ for $n = 0.1$, 
whereas their pattern remains the same as $n$ is increased from 
$0.01$ to $0.1$ for $\hl = 0.1$. There is an overall enhancement 
of $\d\vph$ though, at all epochs (regardless of the oscillations), 
in both the cases. In fact, the enhancement is much larger in the 
former case than in the latter, which is unlike what happens for 
$\dmt$. Nevertheless, this is of not much significance, as $\d\vph$ 
is always subdued (compared to $\dmt$) at small scales, due to its 
unit sound speed ($c_s^2 = 1$), which makes the size of its sound 
horizon of the order of the Hubble radius $\Hp^{-1}$ at the present 
epoch ($N = 0$). However, as shown above, the $\vph$-coupling with 
matter can lead to a fairly considerable effect of $\d\vph$ on the 
evolution of $\dmt$, surmounting the latter's scale-dependence by 
quite an extent, in the deep sub-horizon regime.

\section{Analysis of the matter perturbation growth} \label{sec:Growth}

Let us now examine closely, and more systematically, the evolution 
of the matter density contrast $\dmt$, by referring back to the 
equation (\ref{dmt-k}) derived in the previous section. From the 
subsequent equations therein, viz. (\ref{dphi-k})\,-\,(\ref{F-k}), 
it is clear that the field perturbation $\d\vph$ and the Bardeen 
potential $\F$ are both proportional to $\hl^2$, and therefore 
suppressed immensely at the deep sub-horizon scales ($\hl \ll 1$): 
\bea 
&& \d \vph \,\simeq\, 3 n\, \hl^2 \, \Om \dmt \,, \label{phi-df}\\
&& \F \,\simeq\, \fr{3 \hl^2} 2\, \Om \dmt \,. \label{F-df}  
\eea 
So, from the observational perspective, since the late-time growth 
of the LSS is well within the horizon, it suffices to limit one's 
attention to the evolution of $\dmt$ only. However, the crucial
point to note here is that $\F$ and $\d\vph$, inspite of their 
small scale suppression, may not have a negligible effect on the 
$\dmt$ evolution. This is evident from the rightmost term of 
Eq.\,(\ref{dmt-k}), given by $\le(\F +\, n\, \d\vph\ri)$, modulo 
the factor $\hl^{-2}$ which gets neutralized by the $\hl^2$ 
appearing on the right hand sides of Eqs.\,(\ref{phi-df}) and 
(\ref{F-df}). These equations further imply a practically 
scale-independent evolution of $\dmt$, satisfying the following 
approximated form of Eq.\,(\ref{dmt-k}), for $\hl \ll 1$:
\be \label{DPE}
\dmt_{,_{NN}} + \le[2 \le(1 - 2 n^2\ri) -\, \fr{3 \Om} 2\ri] 
\dmt_{,_N} =\, \fr{3 \le(1 + 2 n^2\ri)} 2 \, \Om \dmt \,,
\ee  
which can be converted to the first-order differential equation 
\be \label{GFE}
f_{,_N} +\, f^2 + \le[2 \le(1 - 2 n^2\ri) -\, \fr{3 \Om} 2\ri] f
=\, \fr{3 \le(1 + 2 n^2\ri)} 2 \, \Om \, ,
\ee
where $\, f(N) := \big[\ln{\dmt} (N)\big]_{,_N} \,$ is known as 
the {\em growth factor}
\cite{GAC-mp,BGNP-NS,CFGPS-Ggi,KD-Rg,TGMP-mpfR}.

\subsection{Growth factor parametrization} \label{sec:Growth-param}

Setting now, certain fiducial values of the matter density 
parameter at the present epoch, $\Omp$, and the coupling 
parameter $n$, we solve the equations (\ref{DPE}) and (\ref{GFE}) 
numerically, to determine the variation of $\dmt$ and $f$ with the 
redshift $z \equiv e^{-N} - 1 $. 
%
\begin{figure}[h]
\centering
\subfloat[{\footnotesize $\dmt$ and $f$ for a fixed $\Omp$ and 
different $n$}]
{\includegraphics[scale=0.6]{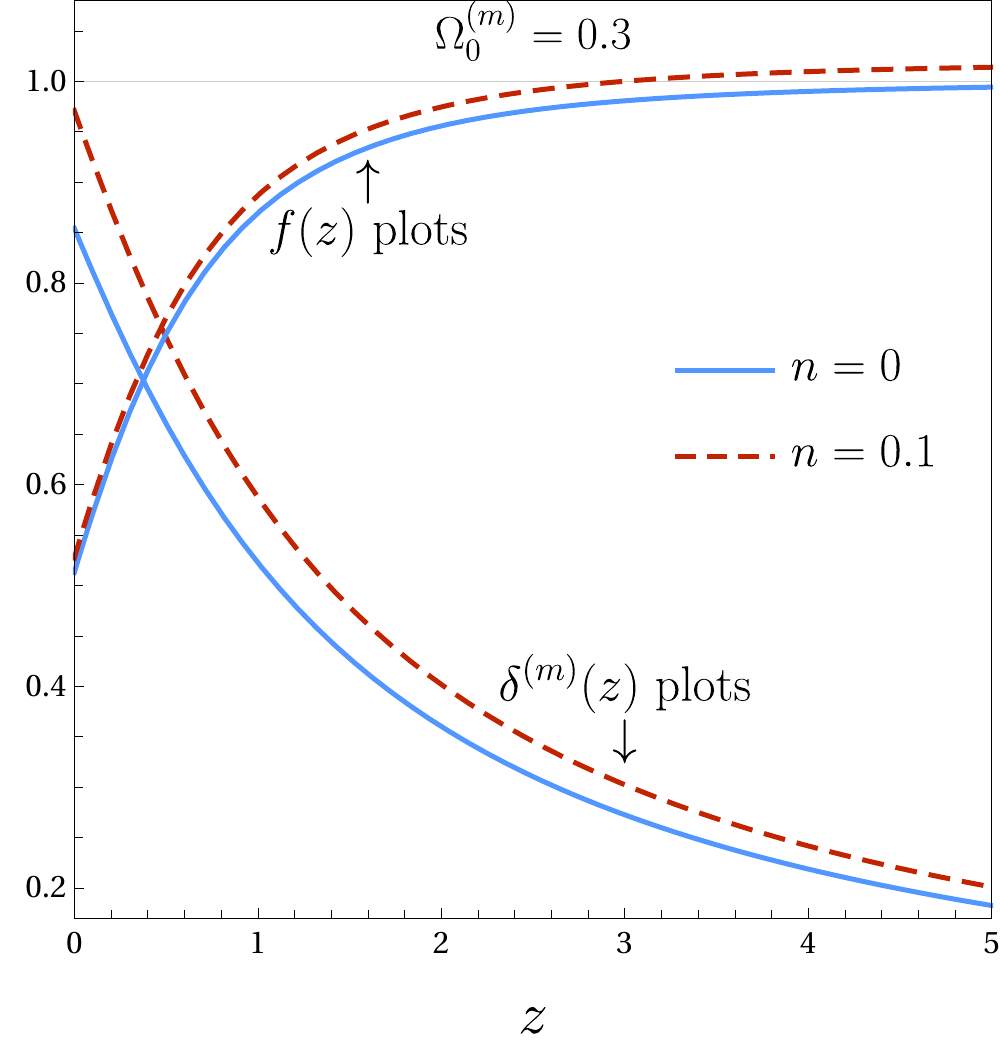}} \quad
\subfloat[{\footnotesize $\dmt$ and $f$ for a fixed $n$ and 
different $\Omp$}]
{\includegraphics[scale=0.6]{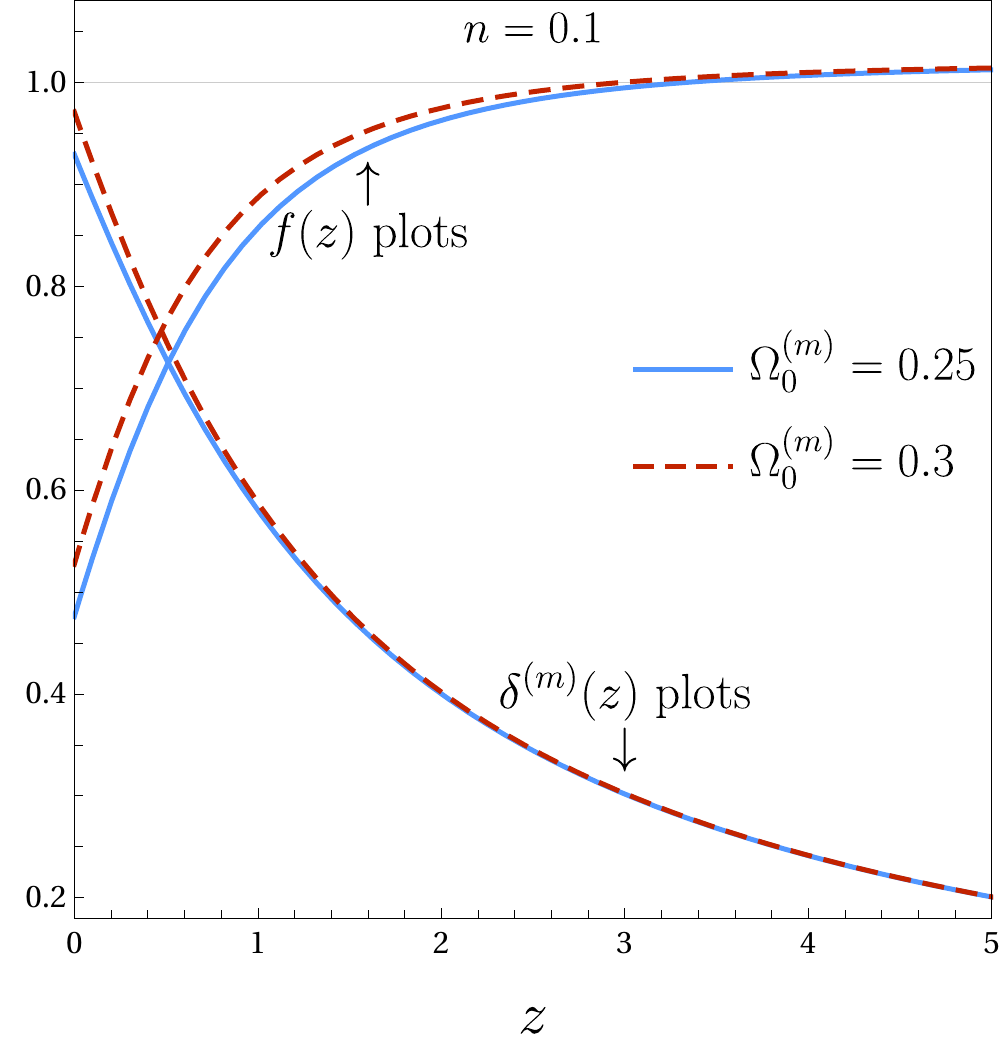}} 
\caption{\footnotesize Matter density contrast $\dmt(z)$ and growth 
factor $f(z)$ in the redshift range $z \in [0,5]$, for (a) fixed 
$\Omp = 0.3$ and different $n = 0, 0.1$, and (b) fixed $n = 0.1$ 
and different $\Omp = 0.25, 0.3$.}
\label{fig:dmf_Om_n}
\end{figure}
%
Figs.\,\ref{fig:dmf_Om_n}\,(a) and (b) show the corresponding plots 
in the redshift range $z \in [0,5]$, respectively, for $\Omp = 0.3$ 
(fixed) and $n = 0, 0.1$, and for $n = 0.1$ (fixed) and $\Omp = 0.25, 
0.3$. 

As we see, with decreasing $z$, both $\dmt(z)$ and its growth rate 
get boosted for a non-zero $n$ and for an increased $\Omp$ in the 
respective cases. However, the same does not hold for $f(z)$ and 
its decay rate correspondingly. While $f(z)$ gets enhanced in both 
the cases, its decay rate gets boosted in the former and diminished 
in the latter. Note further that $f(z)$ exceeds unity at high 
redshifts, when $n$ is non-vanishing. This is really of importance, 
in the sense that it demands one to go beyond the standard 
prescription followed in the matter pertubation growth analysis, 
in presence the DEM interaction. To be more specific, the well-known 
(and widely used) growth factor parametrization ansatz $\, f(N) = 
[\Om(N)]^{\c(N)} \,$, with $\c(N)$ dubbed as the {\em growth index}, 
would only suit the non-interacting scenarios which forbid to any 
breach of the $f = 1$ barrier. One is therefore required to look for 
a modified or alternative parametrization in presence of the DEM 
interaction (i.e. for $n \neq 0$). The following is what we propose, 
by noting that the pre-factor $(1 + 2 n^2)$ on the right hand side 
of Eq.\,(\ref{GFE}) plays the key role in allowing $f$ to exceed 
unity at large $z$ (or high $|N|$), whence $\, \Om \rightarrow 1$:
\be \label{GF}
f(N) \,= \le(1 +\, 2 n^2\ri) \le[\Om(N)\ri]^{\c(N)} \,. 
\ee
%
%
%
%
So the task now boils down to asserting the functional form of the 
growth index $\c(N)$, in an appropriate way, which we discuss below.

\subsection{Growth index fitting} \label{sec:Growth-index}

In the literature, there has been a plethora of functional forms of 
$\c(N)$ considered from various standpoints
\cite{koiv-grow,PA-grow,PG-grow,GP-grow,WYF-grwpara,PAB-grwpara,
BBS-grwpara,BP-grwindx,SBM-grwindx,bat-grwindx,MBMDR-grwindx,
PSG-grwindx,BA-grwindx,HKV-modsel,SS-MSTintDE}.
However, usually they are either suitable for the non-interacting DE 
models only, or valid only at some particular cosmological regimes. 
For instance, the linear order Taylor expansion of $\c(N)$ about 
$z = 0$ remains valid only upto very low redshifts, whereas that 
about $\le[1 - \Om\ri]$ holds good only in the deep matter dominated 
era. Hence, taking the direct route, we substitute Eq.\,(\ref{GF}) 
in Eq.\,(\ref{GFE}) and obtain the following functional fit of the 
solution of the resulting equation numerically, in terms of parameters 
$n$ and $\Omp$: 
\be \label{GI-fit}
\c(N) \,= \le[0.3994 \le(\Omp\ri)^{-0.08} -\, 0.9844 \,n^2\ri] 
\big(1 + e^N\big)^{0.3345} \,.
\ee
This has the obvious advantage of not requiring one to determine first 
the observational constraints on the Taylor coefficients of $\c(N)$, 
while carrying out the statistical estimations of $n$ and $\Omp$, as 
well as the other parameters involved in the analysis, such as the RSD 
parameter $\sep$ and the reduced Hubble constant $h$. 

Note also that in the $\L$CDM limit ($n \to 0$), and for $\Omp = 0.3$, 
the above fit (\ref{GI-fit}) gives $\c_{_0} \equiv \c\vert_{_{N=0}} = 
0.553$, which is very close to the theoretically predicted value $6/11 
\simeq 0.545$ (for $\L$CDM)\footnote{This is actually the value one 
gets by either assuming $\c$ to be a constant, or in a scenario 
(typically the $\L$CDM) in which the sub-leading terms in the Taylor 
expansion of $\c$ about $1-\Om$ happen to be negligible, compared to 
the leading term $\c_{_0}$.}
\cite{WS-intDE}.
%
%
\begin{figure}[h]
\centering
\subfloat[{\footnotesize $E_f(z)$ for a fixed $\Omp$ and different $n$}]{\includegraphics[scale=0.535]{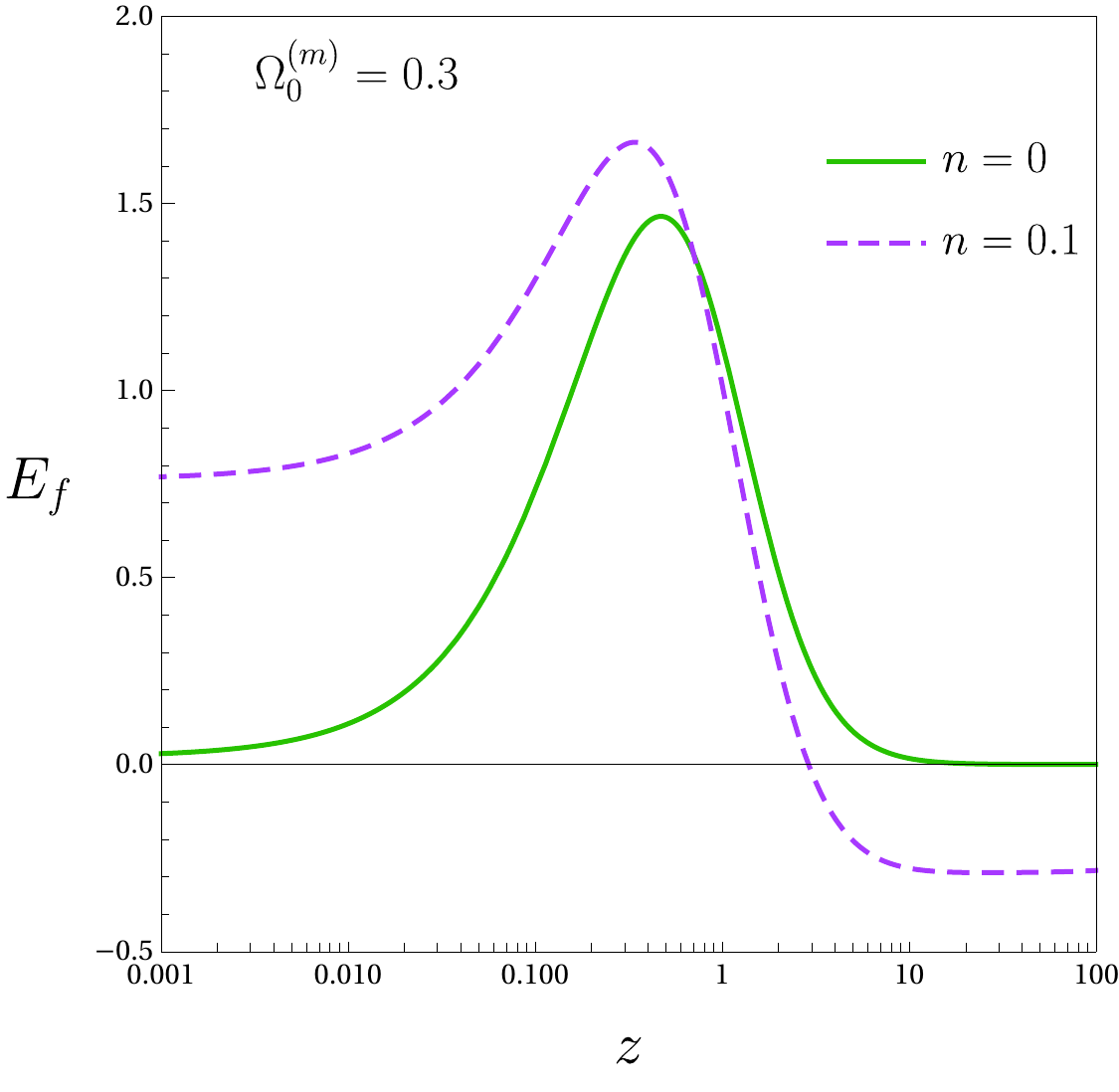}} \quad
\subfloat[{\footnotesize $E_f(z)$ for a fixed $n$ and different $\Omp$}]{\includegraphics[scale=0.535]{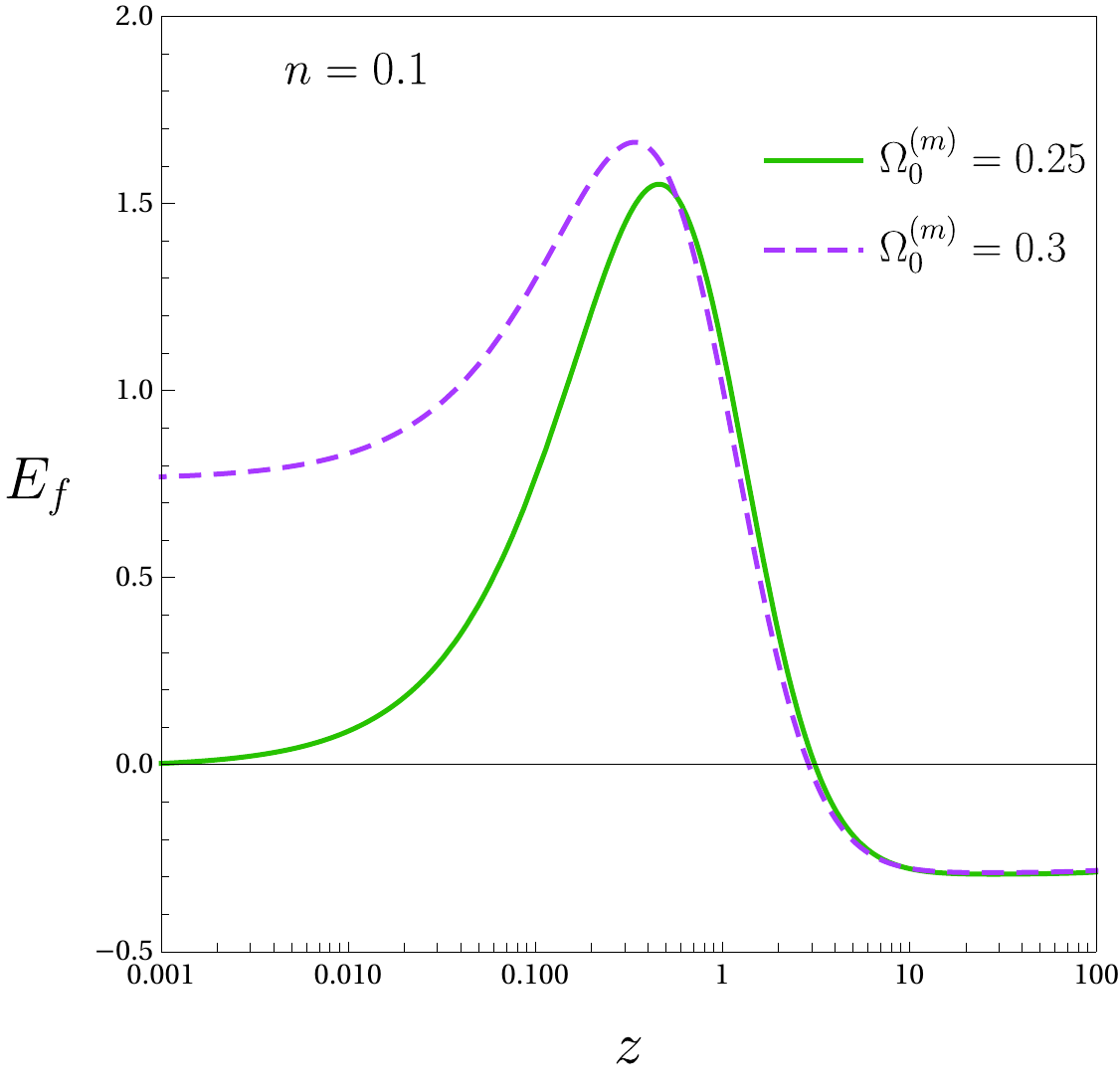}}
\caption{\footnotesize Percentage error $E_f(z)$ in the fitted growth 
factor, for (a) fixed $\Omp = 0.3$ and different $n = 0, 0.1$, and 
(b) fixed $n = 0.1$ and different $\Omp = 0.25, 0.3$, in the redshift 
range $z \in [0,100]$.}
\label{fig:Fit_err}
\end{figure}
%
In fact, the fit is fairly accurate overall, as one may see by examining 
the percentage error
\be \label{GI-fit_err}
E_f (z) \,=\, \fr{100 \big[f_{_F} (z) -\, f(z)\big]}{f(z)} \,\,,
\ee
where $\, f_{_F} (z)$ denotes the fitted growth factor obtained by
plugging Eq.\,(\ref{GI-fit}) in Eq.\,(\ref{GF}) and casting the 
result as a function of $z$, and $f(z)$ is the exact growth factor 
obtained by numerically solving Eq.\,(\ref{GF}) in the preceding 
subsection.

Figs.\,\ref{fig:Fit_err}\,(a) and (b) show the variation of $E_f(z)$ in 
the redshift range $z \in [0,100]$, respectively, for $\Omp = 0.3$ 
(fixed) and $n = 0, 0.1$, and for $n = 0.1$ (fixed) and $\Omp = 0.25, 
0.3$. Except for $z \lesssim 1.5$, the error indeed remains quite low 
($\lesssim 0.5\%$), and does not change too much with the alteration of 
the value of $n$ or of $\Omp$ in the respective cases. In particular,
even for $n$ as high as $0.1$, the maximum error is only about $1.65\%$,
which implies that it is quite reasonable to carry out the parametric
estimation relying on this fit.

\section{Parametric estimation using the growth data} \label{sec:Param-est}

Let us now proceed to carry out the statistical parametric estimation
using the Metropolis-Hastings algorithm for MCMC. 
Our interest is in the RSD observations which measure the quantity
\cite{KPS-brs,NPP-tens,DL-tDE,HSK-mic,KP-tens,ABN-naDEp,BNP-fR}
\be \label{fs8}
\big[f \se\big](z) \,=\, f(z) \, \sep \, \fr{\dmt(z)}{\dmp} \,, 
\ee
where $\, \sep \equiv \se\big\vert_{z=0}\,$ and $\, \dmp \equiv 
\dmt\big\vert_{z=0}\,$. 
Therefore, using the parametrization (\ref{GF}) of the growth factor, and 
its relation to the matter density contrast, we express Eq.\,(\ref{fs8}) 
in the following form (as a function of $N$): 
\bea \label{fs8N}
\big[f \se\big](N) &=&\! \le(1 +\, 2n^2\ri) \le[\Om(N)\ri]^{\c(N)} 
\sep \nn\\
&& \qquad \times\, \exp\le[\le(1 +\, 2n^2\ri) \int_0^ N  dN 
\le[\Om(N)\ri]^{\c(N)}\ri] \,,
\eea
where $\c(N)$ is given by its numerical fit (\ref{GI-fit}) in terms of $n$ 
and $\Omp$.

Now, the statistical analysis is essentially the standard minimization of
\be {\label{chisq}}
\chi^2 :=\, \sum _{i,j} \Big[A_{obs}(z_i) -\, A_{th}(z_i)\Big] \cdot 
C_{ij}^{-1} \cdot \Big[A_{obs}(z_i) -\, A_{th}(z_i)\Big] \,, 
\ee
where $A_{obs}(z_i)$ denotes the observed $f \se$ at a particular redshift
$z_i$, at which the theoretical value $A_{th}$ of $f \se$ can be obtained 
from Eq.\,(\ref{fs8N}), and $C_{ij}$
%
%
is the covariance matrix for three WiggleZ data points 
\cite{KP-tens}
in the refined RSD subsample, viz. the Growth GOLD dataset
\cite{SNS-GR}, 
we are using in this paper.
%
\begin{figure}[h]
\centering
\subfloat[{\footnotesize $2$-D posterior distribution for GOLD dataset}]
{\includegraphics[scale=0.36]{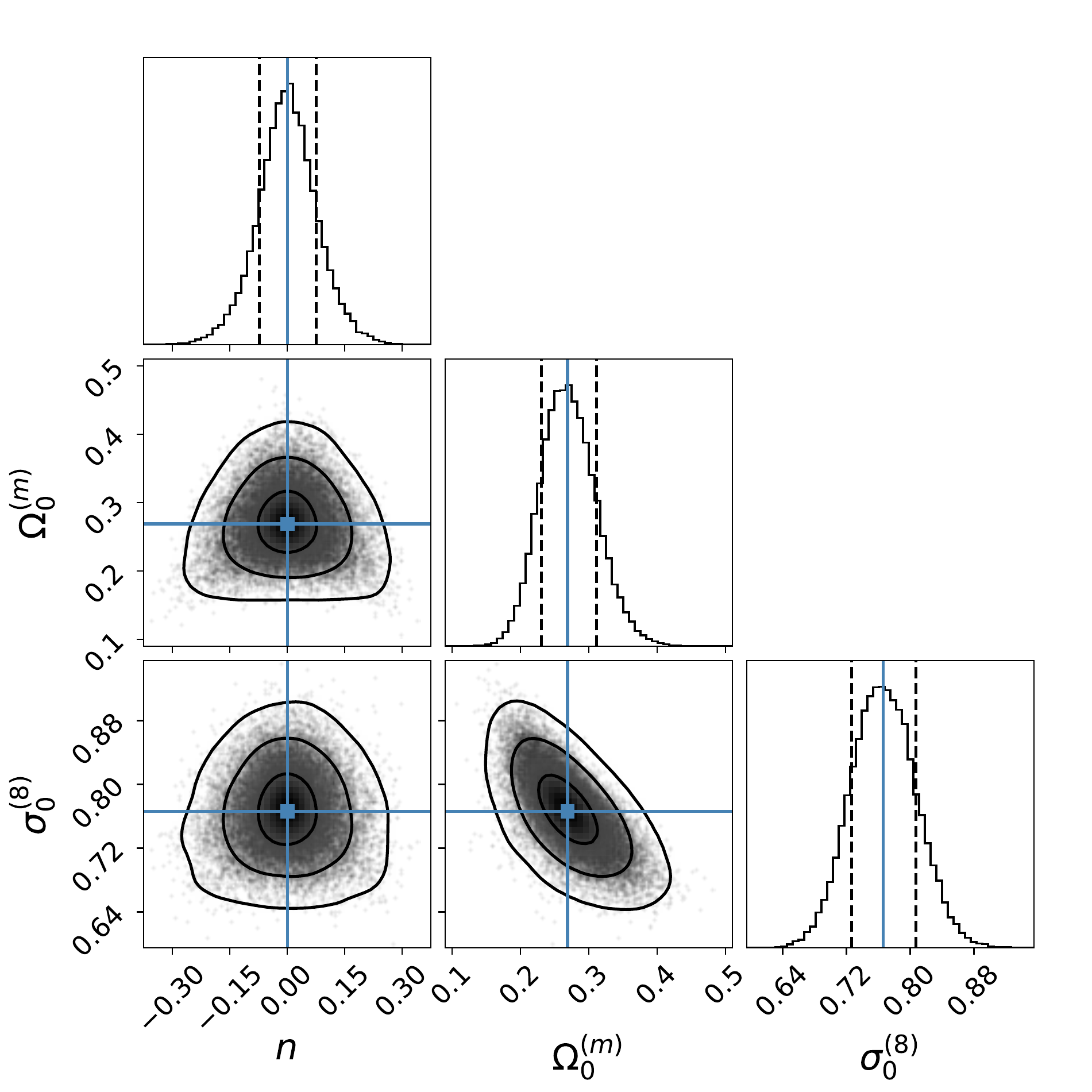}} \\
\subfloat[{\footnotesize $2$-D posterior distribution for GOLD$+ H(z)$ 
dataset}]
{\includegraphics[scale=0.36]{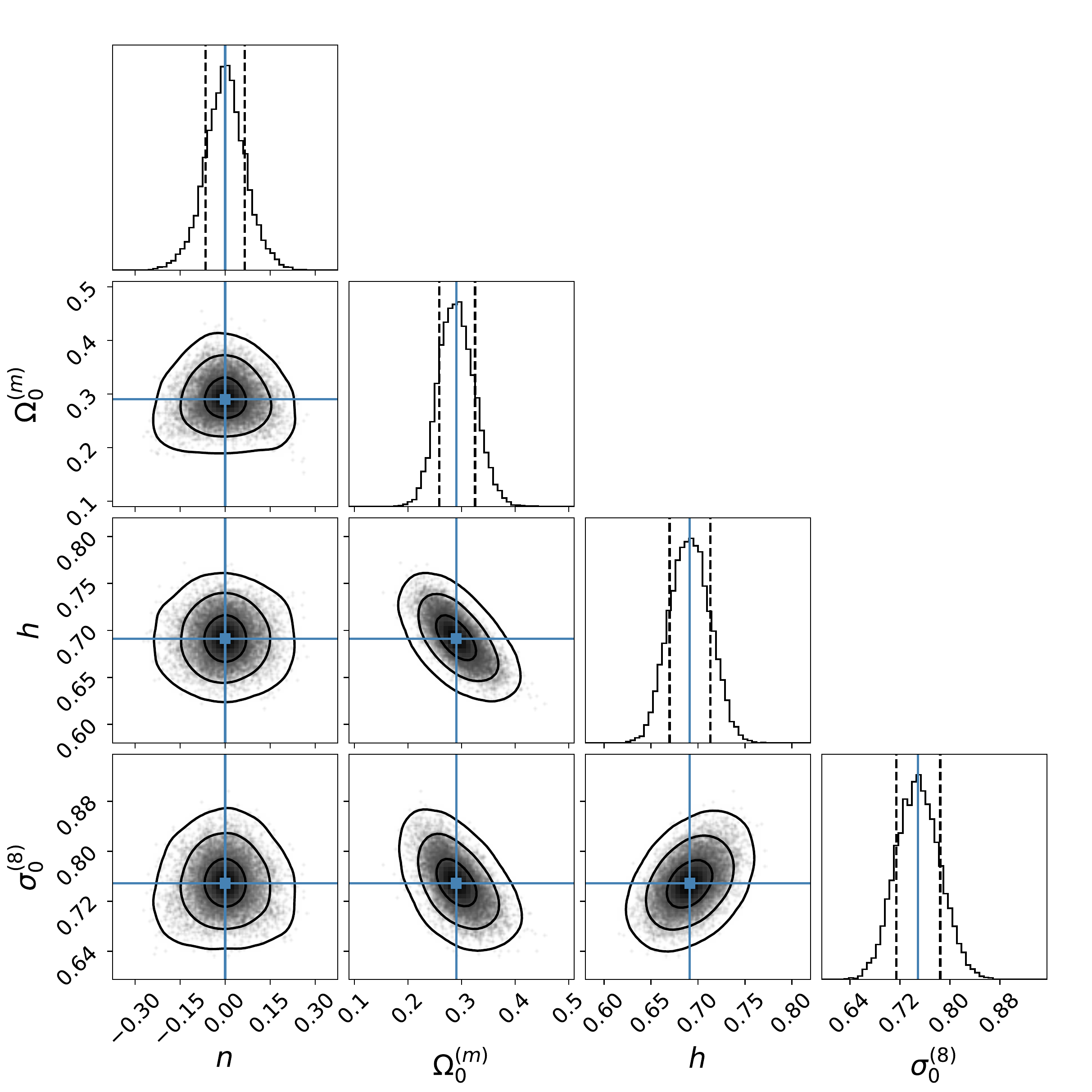}}
\caption{\footnotesize Upto $3\s$ contour levels for (a) the GOLD dataset, 
and (b) its combination with the observational Hubble dataset. The dashed 
lined on the histograms indicate the $1\s$ levels, whereas the solid lines 
on each plot denote the median best-fits of the respective parameters. }
\label{fig:contours}
\end{figure}
%
There are three parameters, viz. $n$, $\Omp$ and $\sep$, that are to be 
constrained in this analysis. An additional parameter, viz. the reduced 
Hubble constant $h$, appears (and needs to be constrained as well) in 
the analysis with the GOLD dataset and the observational Hubble ($H(z)$) 
dataset
\cite{RDR-bao} 
combined. 

Setting the prior ranges
\bea \label{priors}
&&n \in [-0.35,0.35] \,, \qquad \Omp \in [0.1,0.5] \,, \nn\\ 
&&\sep \in [0.6,0.95] \,, \qquad h \in [0.55,0.85] \,,
\eea
we carry out the analysis, with the GOLD and the GOLD+$H(z)$ datasets, 
and determine the allowed parametric domains upto $3\s$. 
%
\begin{table*}[ht]
\tbl{Estimates of the parameters $\Omp$, $\sep$, $h$ and $n$, upto $1\s$ 
limits, alongwith the minimized $\chi^2$, for the GOLD and GOLD+$H(z)$ 
datasets.}
{
\centering
\renewcommand{\arraystretch}{2}
{
\begin{tabular}{||c||c|c|c|c||c||}
\hline
& \multicolumn{4}{c||}{Parametric estimates} & \\
Observational  & \multicolumn{4}{c||}{(best fit and $1\s$ limits)} & 
$\chi^2_{min}$ \\
\cline{2-5}
%
Datasets & $\Omp$ & $\sep$ & $h$ & $|n|$ &  \\
\hline\hline
GOLD & $ 0.269^{+0.042}_{-0.038}$ & $0.766^{+0.041}_{-0.040}$ & -- & 
$ < 0.074$ & $13.152$ \\
\hline
GOLD+$H(z)$ & $0.290^{+0.035}_{-0.032}$ & $0.748^{+0.036}_{-0.034}$
& $0.691^{+0.022}_{-0.021}$ & $ < 0.065$ & $30.662$ \\
\hline\hline
\end{tabular}
}
\label{Est-tab}
}
\end{table*} 
%
Figs.\,\ref{fig:contours}\,(a) and (b) show the same, for the respective 
cases, whereas the Table\,\ref{Est-tab} shows the corresponding parametric 
estimates, upto $1\s$ confidence limits, as well as the minimized values 
of $\chi^2$. 

The parameters are more tightly constrained by the GOLD+$H(z)$ 
dataset, than by the GOLD dataset, which is of course expected with the 
increased number of uncorrelated data points in the former. However, note 
that even for the GOLD+$H(z)$ dataset, the marginalized $\Omp$ shown in
Table\,\ref{Est-tab} is not much deviated from that for $\L$CDM, such as 
$\, 0.3111 \pm 0.0056 \,$ obtained from the 
Planck 2018 TT,TE,EE+lowE+Lensing+BAO combined analysis
\cite{Planck18-CP}.
It is also worth pointing out here that the best fit value of the DEM 
coupling parameter $n$ turns out to be quite insignificant (about 
three orders of magnitude below unity). This is the reason why in 
the Table\,\ref{Est-tab} we have only quoted the $1\s$ limit of $n$, 
which is quite small as well. All these imply only a very mild effect 
of such a coupling on the $\L$CDM background evolution, attained in 
the limit $n \to 0$ of the solution (\ref{Hub-sol}). 

On the whole, therefore, we infer that the background dynamics of 
the effective DE component, described by Eq.\,(\ref{Hub-sol}), is 
constrained to be very weak by the Growth GOLD subsample, and even 
more so, by the combination of the latter with the $H(z)$ data. 
This corroborates to not only the general indications of the 
cosmological observations, but also the solar system constraints 
on the scalar-tensor theories. More specifically, the smallness of 
$n$, which is the root cause of such a weak DE dynamics, tallies 
with the very large upper bound on the effective Brans-Dicke 
parameter $\fw$ which is linear in $n^{-2}$ (see Eq.\,(\ref{A-redef}) 
in the Appendix)
\cite{TUMTY-ST,Acq-BDbound,AS-BDbound,CW-BDbound,Als-BDbound}.

Nevertheless, the consequences of the DEM interaction in the cosmological
perturbative spectrum do not seem to be that mild, from the amount of 
deviation of the derived estimate of the growth index at the present 
epoch, $\c_{_0} \equiv \c\big\vert_{z=0}$, from the corresponding 
predicted value $6/11 \simeq 0.545$ for $\L$CDM
\cite{WS-intDE}.
%
\begin{figure}[h]
\centering
\subfloat[{Estimated $\c(z)$ for GOLD}]
{\includegraphics[scale=0.44]{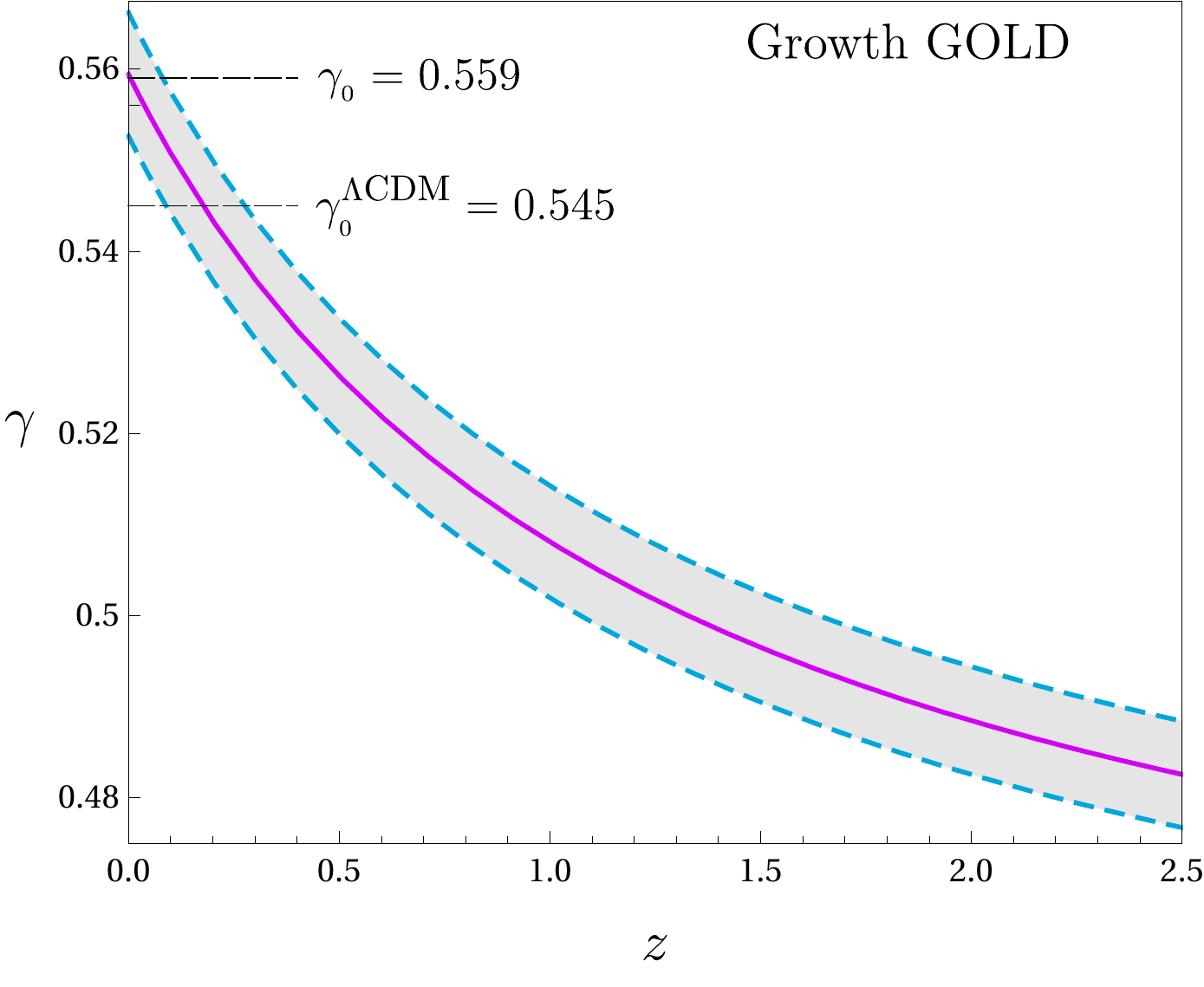}} ~
\subfloat[{Estimated $\c(z)$ for GOLD+$H(z)$}]
{\includegraphics[scale=0.44]{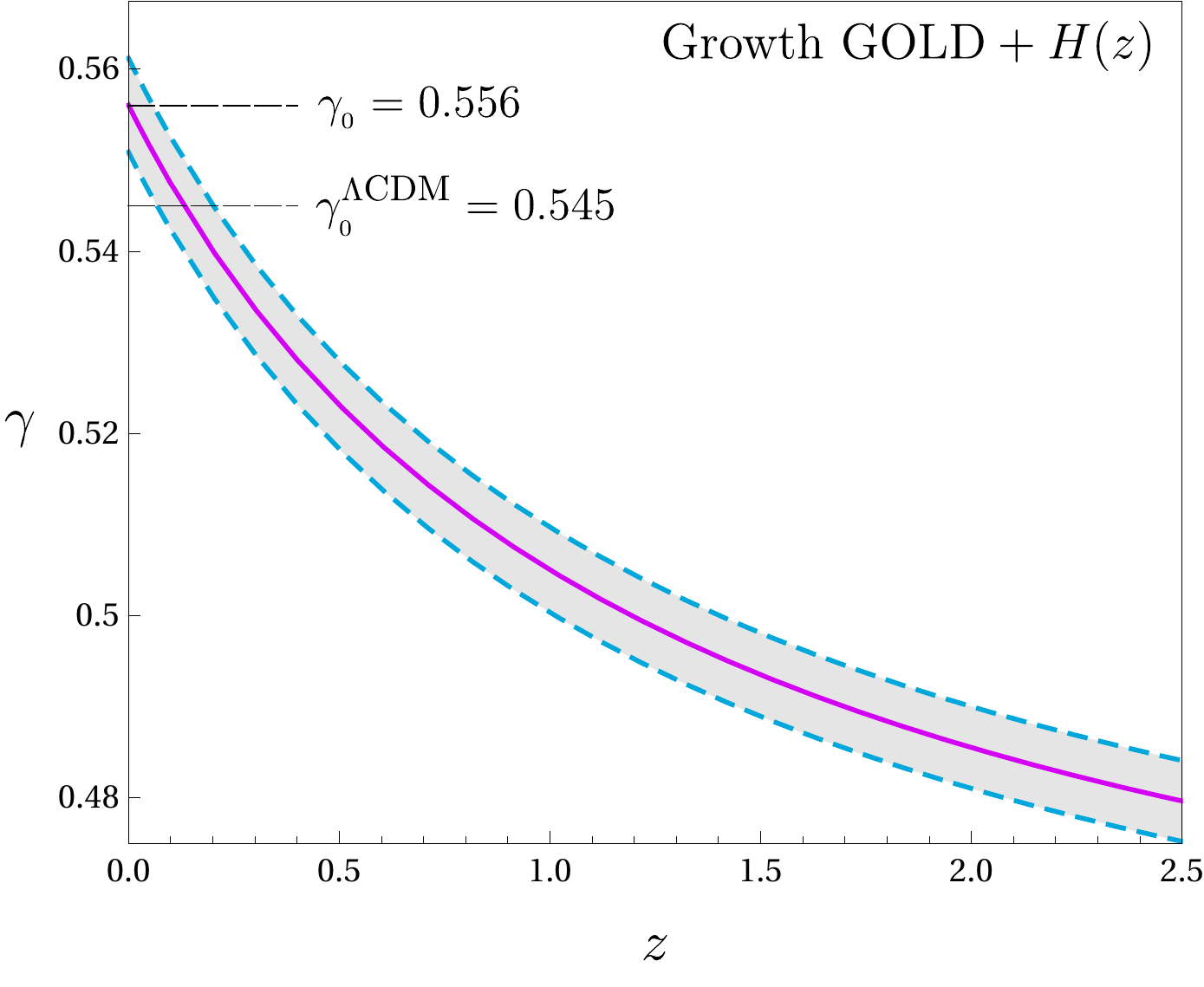}} 
\caption{\footnotesize Evolutionary profile of the growth index $\c(z)$
estimated using (a) the GOLD dataset and (b) the GOLD+$H(z)$ dataset. 
The solid lines in the plots denote the best-fit and the dashed lines 
correspond to the respective $1\s$ margins. }
\label{fig-gmama-est}
\end{figure}
%
Figs.\,\ref{fig-gmama-est}\,(a) and (b) show such a deviation in each
of the plotted evolution profiles of the best fit $\c(z)$ and the 
corresponding $1\s$ margins derived using the parametric estimations 
with the GOLD dataset and with the GOLD+$H(z)$ dataset, respectively. 
Specifically, the estimates are $\,\c_{_0} = 0.559 \pm 0.007$ for 
GOLD and $\,\c_{_0} = 0.556 \pm 0.005$ for GOLD+$H(z)$, which imply
that the predicted $\L$CDM value $0.545$ is not within the $1\s$ 
error limits in either case, even though the coupling parameter $n$ 
is estimated to be very small. 

\begin{figure}[h]
\centering
\subfloat[{Estimated $f(z)$ for GOLD}]
{\includegraphics[scale=0.44]{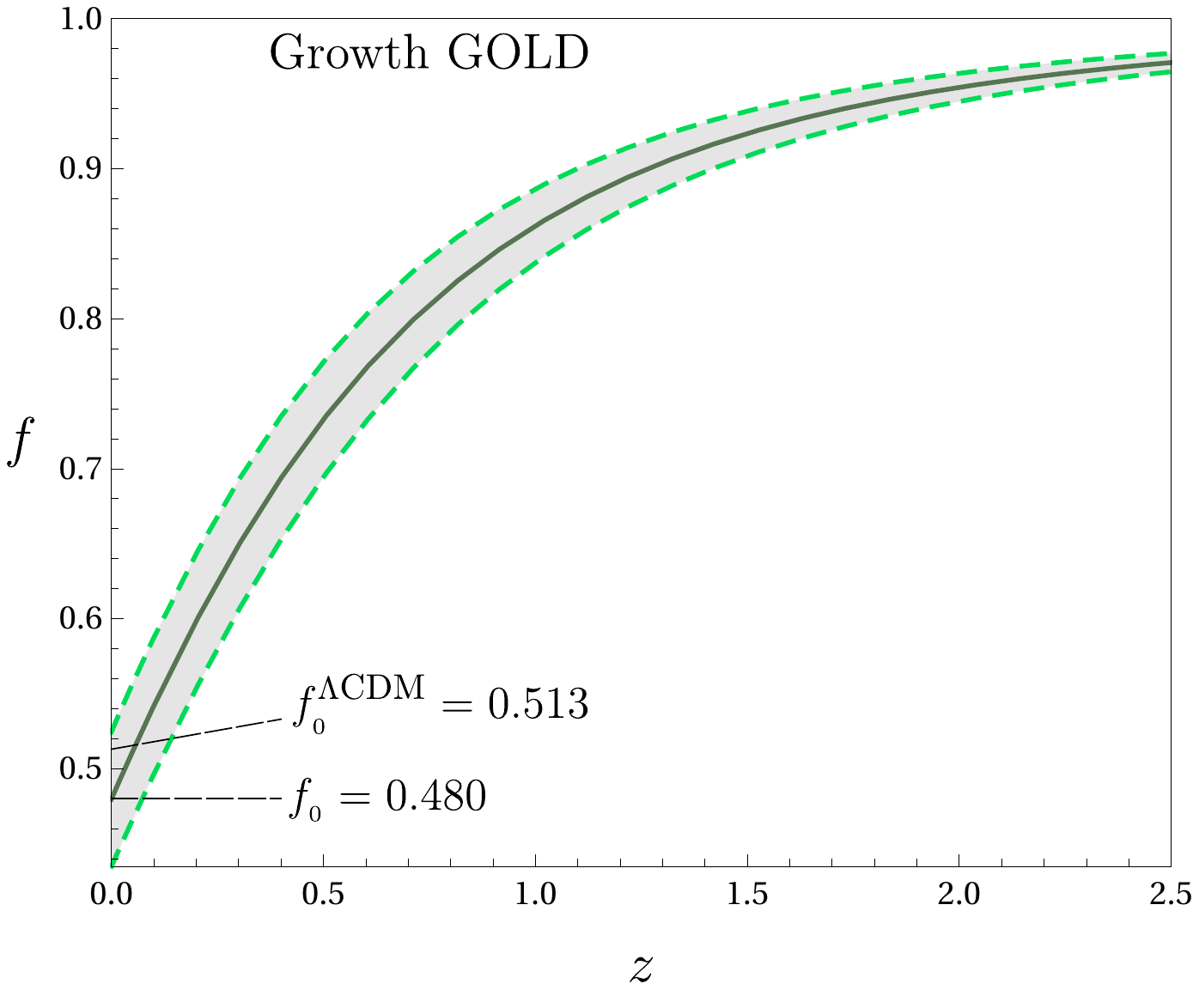}} ~
\subfloat[{Estimated $f(z)$ for GOLD+$H(z)$}]
{\includegraphics[scale=0.44]{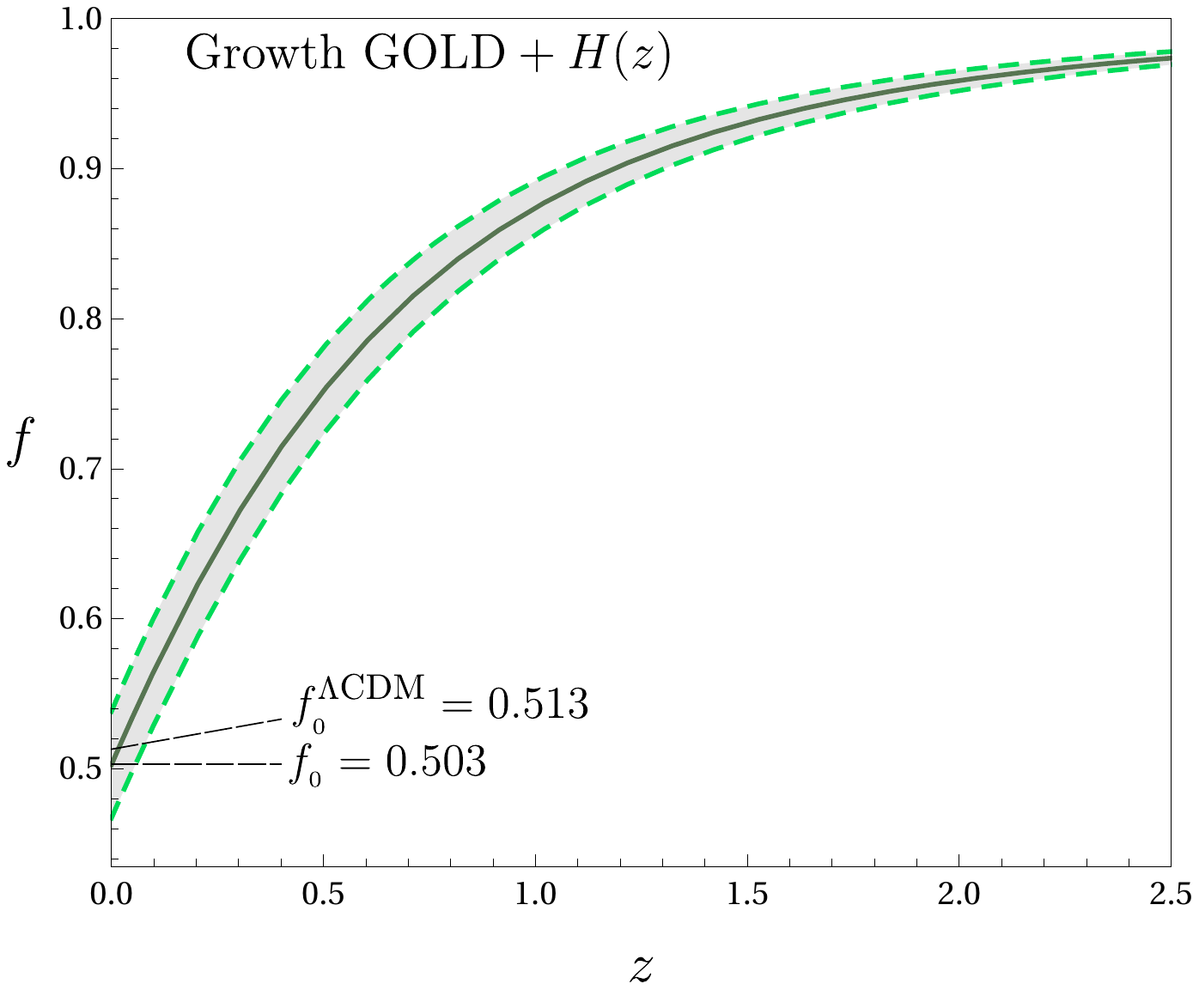}} 
\caption{\footnotesize Evolutionary profile of the growth factor $f(z)$
estimated using (a) the GOLD dataset and (b) the GOLD+$H(z)$ dataset. 
The solid lines in the plots denote the best-fit and the dashed lines 
correspond to the respective $1\s$ margins. }
\label{fig-f-est}
\end{figure}

\begin{figure}[h]
\centering
\subfloat[{Estimated $f\se(z)$ for GOLD}]
{\includegraphics[scale=0.44]{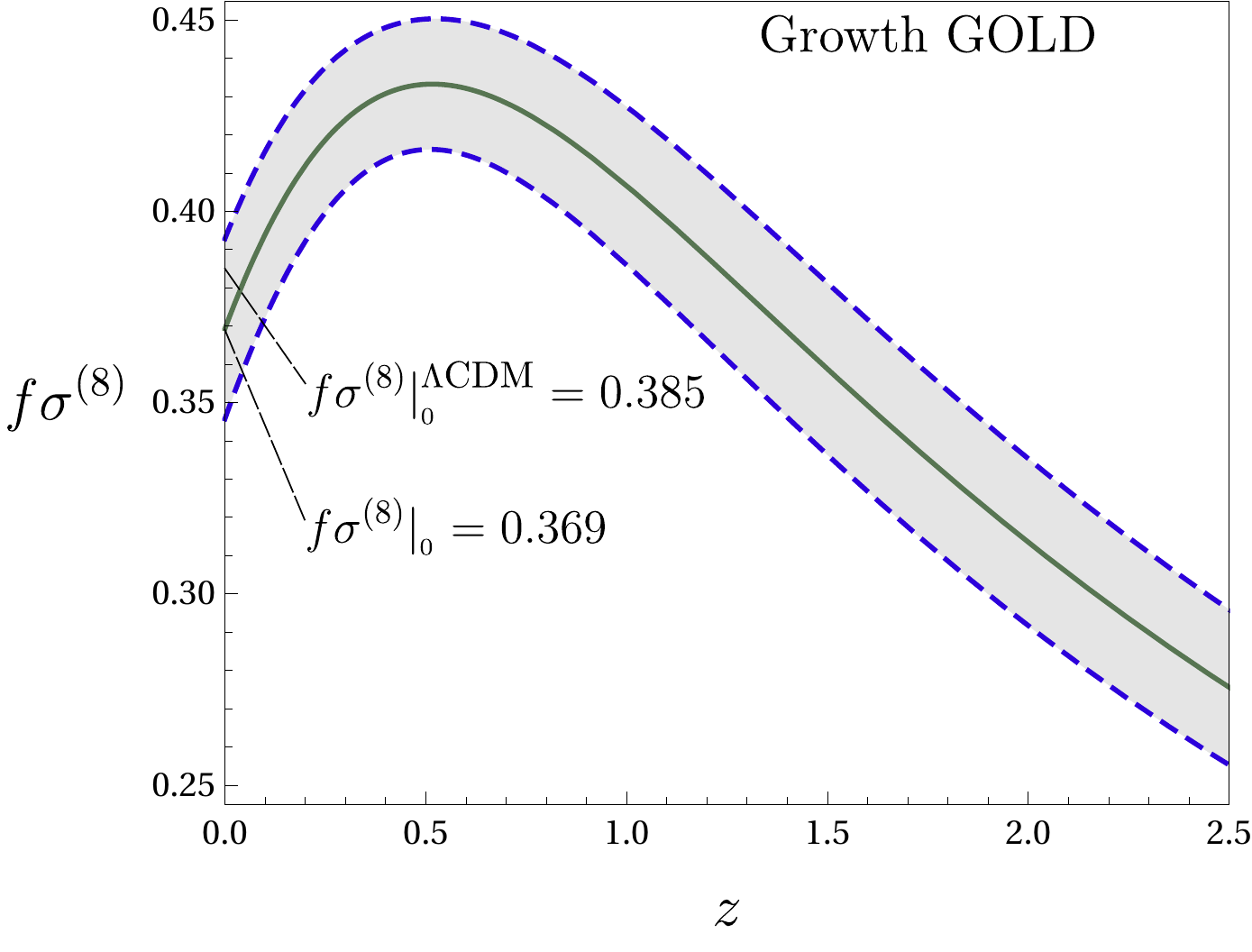}} ~
\subfloat[{Estimated $f\se(z)$ for GOLD+$H(z)$}]
{\includegraphics[scale=0.44]{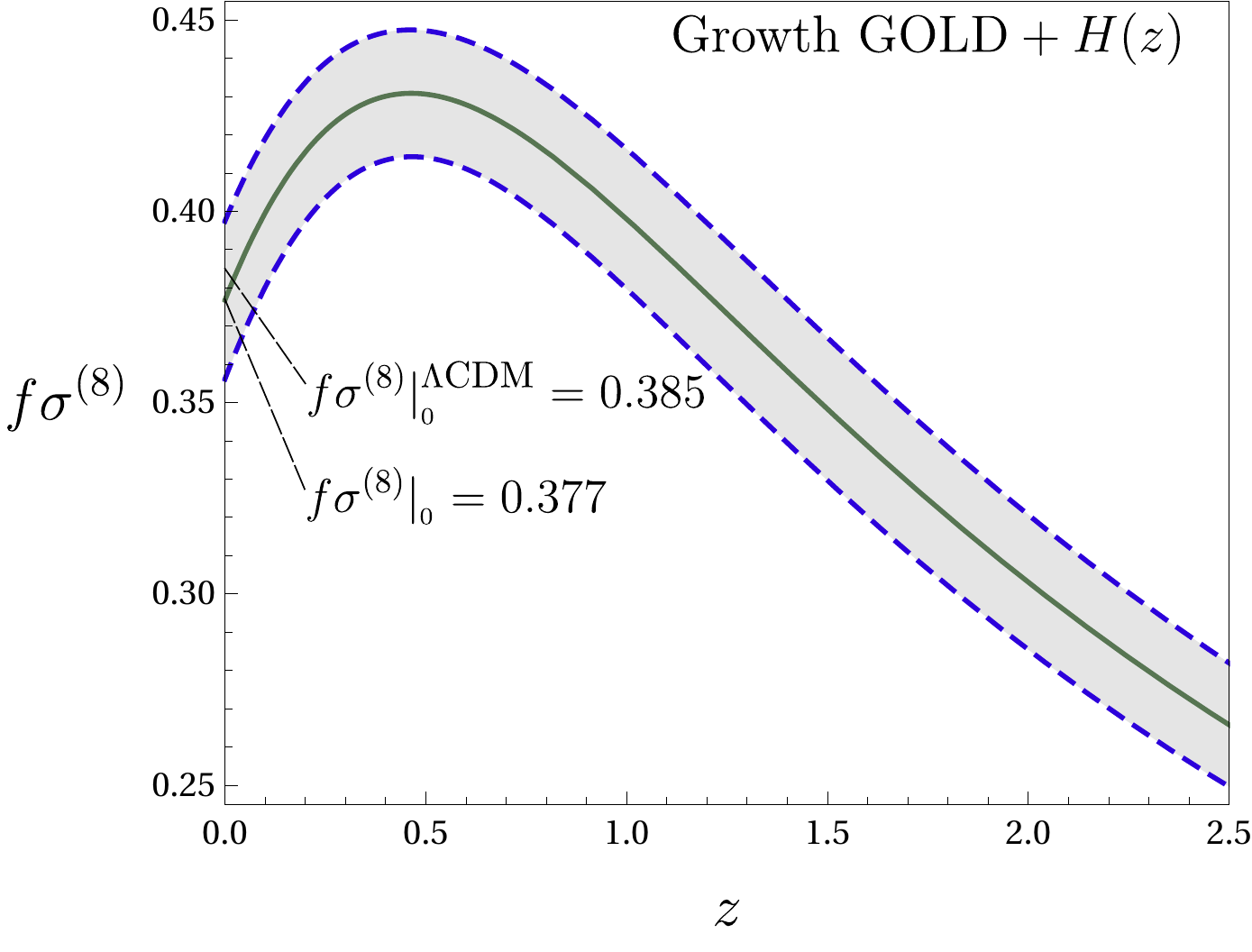}} 
\caption{\footnotesize Evolutionary profile of $f \se(z)$
estimated using (a) the GOLD dataset and (b) the GOLD+$H(z)$ dataset. 
The solid lines in the plots denote the best-fit and the dashed lines 
correspond to the respective $1\s$ margins. }
\label{fig-fs8-est}
\end{figure}

One may note that these estimates of $\c_{_0}$ are consistent with 
those obtained for many dynamical DE models in the literature, the 
non-interacting ones as well as those in which the DE interacts with 
matter (albeit, mostly phenomenologically)
\cite{koiv-grow,PA-grow,PG-grow,GP-grow,amen-pert,SHCK-grow}.
However, it may cause an inherent deception if we infer a very 
significant change over $\L$CDM just from what shown in 
Figs.\,\ref{fig-gmama-est}\,(a) and (b). 
%
%
In particular, unlike the $\L$CDM value of $\c_{_0}$, the 
corresponding $f(z)$ and $f(z)\se(z)$ values at the present epoch 
do not fall beyond the $1\s$ confidence limit of our estimates for 
both the datasets, as shown in Figs.\,\ref{fig-f-est}\,(a),(b) 
and in Figs.\,\ref{fig-fs8-est}\,(a),(b), respectively. 
%
%
%
The contrary aspect of $\c_{_0}$ is actually due to the fact that 
it is not an independent parameter in our formulation, but is a 
redefined one, the error on which is primarily dependent on the 
estimates of the parameters $n$ and $\Omp$. Hence, the error-propagation 
method leads to a very small $1\s$ interval of $\c_{_0}$, of the order 
of $10^{-3}$. On the other hand, the $f(z)$ and $f(z)\se(z)$ plots 
in Figs.\,\ref{fig-f-est}\,(a),(b) and \ref{fig-fs8-est}\,(a),(b) 
are obtained by using our growth factor ansatz (\ref{GF}) for the 
range $z \in [0,2.5]$, upto $1\s$, for either dataset. The $1\s$ 
uncertainties in the respective values at the present epoch, viz. 
$f_{_0}$ and $f\se|_{_0}$, turn out to be of the order of $10^{-2}$, 
within which the $f_{_0}$ and $f\se|_{_0}$ values corresponding to 
$\L$CDM get easily accommodated.  
This in fact confirms that the ansatz (\ref{GF}) is consistent with 
the $\L$CDM model as well. 

For a further clarification, let us recall that the fitting function 
(\ref{GI-fit}) involves a linear dependence of the growth index $\c$
on $n^2$. However, since the estimated $n$ is as large as 
$\mathcal{O}(10^{-2})$ upto the $1\s$ limit, we may safely argue 
that the dependence of $\c$ on $n$ alone would not be sufficient 
enough to account for the fall-out of the $\L$CDM value $0.545$ of 
$\c_{_0}$ beyond the estimated $1\s$ level of $\c$. It is actually 
the latter's dependence on $\Omp$ which plays a major role here. If 
suppose this dependence is ignored, i.e. the same $\Omp$ value (say, 
$0.29$) is assumed for all values of $n$, then for $n = 0$ ($\L$CDM) 
the fitting gives $\c_{_0} = 0.5561$, whereas for $n = 0.01$ (say) 
one gets $\c_{_0} = 0.5559$. This clearly shows how crucial is the 
$\Omp$-dependence, in comparison to the $n$-dependence of $\c$.

Let us also point out that one should not get deceived by the maximum 
percentage error ($\sim 1.65\%$) in our fitting, as shown by the 
plots in Figs.\,\ref{fig:Fit_err}\,(a) and (b). Such a deviation is 
due to the fiducial value of $n \,(= 0.1)$ we have assigned while 
obtaining such plots. The actual estimates of $n$ shown in 
Table\,\ref{Est-tab} is far less than that value, even upto the 
$1\s$ limit, for both the datasets we have considered. If we use 
these estimates, then the percentage error in the fit would have a 
much lesser maximum value than $\sim 1.65\%$, and would be negligible 
at the present epoch ($z=0$). 

On the whole, therefore, we may infer that the exclusion of the 
$\L$CDM growth index value $\c_{_0} \simeq 0.545$ from the estimated 
$1\s$ domain of $\c$ is not because of the error in the fitting but 
due mainly to the uncertainty in the parametric estimation of $\Omp$, 
with a supporting role played by the uncertainty in the parameter $n$ 
as well as its covariance with $\Omp$.

\section{Conclusion} \label{sec:concl}

We have thus made a comprehensive study of the evolution of the 
cosmological perturbations in a dark energy-matter (DEM) interacting 
scenario that emerges from a class of scalar-tensor theories, motivated 
from the perspective of a wide range of modified gravity theories. 
While giving a general account on its formulation, we have emphasized 
on the plausible forms of the potential term for the scalar field, that 
would lead to exact solutions of the cosmological equations in the 
standard setup. Resorting to one such form of the potential, which is 
simply a mass term in the (original) Jordan frame, or equivalently an 
exponential term in the Einstein frame, we have discussed the merits 
of the cosmological solution one obtains thereby. One advantage is 
that the DEM coupling strength is measured by just the one parameter 
$n$, which is inversely related to square root of the Brans-Dicke 
parameter $\fw$. The dynamics of the DE, induced by the scalar field 
$\vph$ which the solution describes, is however expected to be quite 
mild, in view of the strong observational favour on $\L$CDM (that is 
attained by switching off $n$). Nevertheless, in spite of the mildness 
of the DE dynamics, the DEM interaction may leave some imprint on the 
pertubative spectrum, while the background cosmology is governed by
the abovementioned solution. This is what we have contemplated on in 
this paper.

Choosing to carry out the perturbative analysis in the well-known 
Newtonian gauge, we have derived the evolution equations for the matter 
density contrast $\dmt(z)$, the growth factor $f(z)$ and the scalar 
field perturbation $\d\vph$. These equations are highly coupled though, 
and an effective decoupling could be achieved only at the deep 
sub-horizon scales, on which we have kept our attention. In fact, 
remember that even after making a specific gauge choice (to eliminate 
the unreal perturbations) one is left to choose the size of the 
horizon, as the perturbations behave differently depending on that 
size. Now, the deep sub-horizon regime is of relevance in the sense 
that $\dmt(z)$ is nearly scale-independent therein, so that one can 
make the most of out of the RSD observations for the LSS formed well 
inside the radius of the horizon. Moreover, in this regime, the 
perturbation in the $\vph$-induced DE, i.e. $\d\vph(z)$, is highly
suppressed compared to the total matter density perturbation 
$\d\rmt(z)$, and hence the onus is completely on $\dmt(z)$.
Nevertheless, the DEM interaction changes the scenario a bit, as 
$\d\vph(z)$ makes an additional, albeit small, contribution to 
$\dmt(z)$, and hence to the growth factor $f(z)$. To be more precise,
the pertubation $d\vph(z)$ is oscillatory at the sub-horizon scales. 
However, the oscillations are about a mean value is proportional to
the DEM coupling parameter $n$. So, for $n \neq 0$ a contribution
would come from the time-averaged $\d\vph(z)$. 

Now, the point to be reckoned here is that $n$ is expected to be quite 
small, as otherwise the background DE dynamics would not be considered 
as mild. Therefore, to what extent would the effect of the overall 
$n$-dependence of $\dmt(z)$ be of significance (if at all), compared to 
the latter's scale-dependence, which is anyway neglected in the deep 
sub-horizon regime? We have addressed this issue by simultaneously 
solving the evolution equations for $\dmt(z)$ and $\d\vph(z)$ 
numerically, for some fiducial settings of $n$ and $\hl$, the 
perturbation scale. As it turned out, the deviation in the $\dmt$ 
evolution profile is quite less for a single order-of-magnitude change 
in $\hl$ with $n$ kept fixed, than for the exact converse. Although 
$n$ and $\hl$ are characteristically different, this gives at least 
a qualitative idea that the coupling ($n$)-dependence is quite 
overwhelming than the scale ($\hl$)-dependence of $\dmt$. In other 
words, the scale-independence of $\dmt$ in the deep sub-horizon regime
does not necessarily mean a negligible effect of the DEM interaction.

While the deep sub-horizon limit allows a scale-independent formulation 
of $\dmt(z)$, one finds the desired decoupling of the evolution 
equations. The equation for the growth factor $f(z)$ can in principle 
be solved by using a suitable parametrization ansatz. As $\dmt(z)$ 
essentially depends on the matter density parameter $\Om(z)$, and so 
does $f(z)$, it is quite justified to consider the parametrization 
$f(z) = [\Om(z)]^{\c(z)} \,$, where $\c(z)$ is referred to as the 
growth index. Such a parametrization is well-known and widely used in 
the literature, as it provides a convenient way of discriminating 
various DE models and checking their viability. Nevertheless, it has
the drawback of not allowing $f(z)$ to exceed unity, and as such
incompatible with the configuration we have in presence of the DEM
interaction. So we have modified this parametrization to $f(z) = 
(1 + 2 n^2) [\Om(z)]^{\c(z)} \,$, and looked to carry on with the 
formulation in terms of the growth index $\c(z)$. However, contrary
to the common practice, we have refrained from resorting to any 
Taylor expansion of $\c(z)$, as it usually remains valid only for 
small range of redshifts or does not take into account the explicit 
dependence on the coupling parameter $n$. Instead, we have taken the 
direct approach, i.e. to substitute the parametrized form of $f(z)$ 
in its evolution equation and solve the latter to obtain a numerical 
fitting function for $\c(z)$ in terms of the parameters $\Omp = 
\Om\big\vert_{z=0}$ and $n$. The fitting function is found to be 
quite accurate even at redshifts $z \simeq 100$ or more. Hence, 
both the issues with the Taylor expanded forms are alleviated.

The allowed parametric spaces are finally determined using the 
Metropolis-Hastings algorithm for MCMC, for two chosen datasets, 
viz. the RSD GOLD subsample and that combined with the observational 
$H(z)$ data. The combined dataset constrains the parameters more 
tightly, however, even for this dataset the marginalized $\Omp$ shows 
only a mild deviation from that for $\L$CDM, as quoted in the Planck 
2018 results
\cite{Planck18-CP},
for example. The estimated $n$, upto $1\s$, is very small ($0.074$ 
and $0.065$ for the GOLD and Gold+$H(z)$ respectively). So the DEM 
coupling, or in fact the DE dynamics, is very weak. Notwithstanding 
the general agreement with the cosmological observations, this 
corroborates to the astrophysical constraints on the scalar-tensor 
theories. However, the overall constraints on the parameters 
(including $\sep$ and $h$, the reduced Hubble constant) do not
indicate that mild an effect of the DEM interaction on the growth
of the matter density contrast $\dmt(z)$. This is noticed from the 
amount of deviation of the growth index at the present epoch, 
$\c_{_0}$, estimated using such constraints, from the corresponding 
predicted value for $\L$CDM
\cite{WS-intDE}.
Nevertheless, such a deviation does not make anything clear on the
individual contributions to $\dmt(z)$, i.e. the one due to the drag
force on the latter, as the coupling makes the background matter 
density drift from its usual dust-like form, and the other due to 
the fluctuations of DE perturbation $\d\vph(z)$ about a non-zero 
mean value proportional to the coupling parameter $n$. A more 
in-depth study is required to assert this. Such a study is presently 
being done in a subsequent work, which we hope to report soon. 

\section*{Acknowledgments}
The work of MKS was supported by the Council of Scientific and Industrial Research (CSIR), Government of India. 

\appendix
\section{DEM interaction in a class of Scalar-Tensor theories} 
\label{app}

The Jordan frame action for scalar-tensor theories is generally expressed,
in presence of matter fields $\{\psi\}$, as 
\bea \label{A-J-ac}
&& \cS =\, \fr 1 2 \int d^4 x \sq{-g} \Big[f(\f) R(g_{\m\n}) -\, \fw(\f)\, 
g^{\m\n} \pa_\m\f \, \pa_\n \f -\, V(\f)\Big] \nn\\ 
&& \qquad \qquad \qquad \qquad 
+ \int d^4 x \sq{-g}\, \cL^{(m)}(g_{\m\n},\psi) \,\,,
\eea
where $g$ is the determinant of the metric $g_{\m\n}$ of the Jordan frame,
$f(\f)$ and $\fw(\f)$ are two differentiable coupling functions of the 
scalar field $\f$, with $V(\f)$ being the corresponding (self-interacting) 
potential, and $\cL^{(m)}(g_{\m\n},\psi)$ is the matter Lagrangian 
density. Note that, a non-trivial $f(\f)$ implies a non-minimal coupling
of $\f$ and the Ricci curvature scalar $R$, to which the matter fields 
$\{\psi\}$ are minimally coupled. Moreover, one can always redefine the 
field $\f$ in a way that $\fw$ is a constant. 

Consider now a class of such theories, characterized by a quadratic 
form of the non-minimal coupling function, i.e. $f(\f) = \f^2$. This has 
equivalent formulations in a wide range of modified gravity theories,
and also of the Brans-Dicke (BD) theory in absence of the potential 
$V(\f)$. In fact, it is easy to see that for $f(\f) = \f^2$, a field 
re-definition $\f^2 = \widetilde{\f}$ renders the free Jordan frame action 
to the BD action expressed in terms of $\widetilde{\f}$. So the constant 
$\fw$ is nothing but the BD parameter.  
%
%

Under the stipulation
\be \label{A-phi0}
\fp \equiv \f(\tp) =\, \k^{-1} \,, \quad \mbox{where} \quad 
\k^2 = 8 \pi G_N \,,
\ee
one has the effective (running) gravitational coupling factor equal to the 
Newton's constant $G_N$ at the present epoch ($t=\tp$).
On the other hand, a conformal transformation $\, g_{\m\n} \to \hg_{\m\n} = 
\k^2 \f^2 g_{\m\n}\,$, followed by another field re-definition 
\be \label{A-redef}
\f :=\, \k^{-1} \, e^{n \k \vph} \,, \quad \mbox{with} \quad 
n = \le(6 +\, \fw\ri)^{-1/2} \,,
\ee
reduces the Jordan frame action to the Einstein frame action in the form
\be \label{A-E-ac}
\hS =\, \int d^4 x \sq{-\hg} \le[\fr{\hR(\hg_{\m\n})}{2\k^2} -\, 
\fr 1 2\, \hg^{\m\n} \pa_\m \vph \, \pa_\n \vph -\, U (\vph)
+\, \hLm(\hg_{\m\n},\vph,\psi)\ri] , 
\ee
where $\hg$ is the determinant of the corresponding (transformed) metric 
$\hg_{\m\n}$, and
\bea 
&& U(\vph) :=\, \fr{e^{-4 \k n \vph}} 2 \, V(\f(\vph)) \,\,,
\label{A-E-Pot} \\
&& \hLm(\hg_{\m\n},\vph,\psi) :=\, e^{-4 \k n \vph} \, 
\cL^{(m)}(\hg_{\m\n},\f(\vph),\psi) \,\,,
\label{A-E-Lm}
\eea
are, respectively, the scalar field potential and the effective matter 
Lagrangian density in the Einstein frame. 
%

Therefore, with the usual definitions of the energy-momentum tensors 
corresponding to matter and the scalar field $\vph$, viz. 
\bea 
\hT_{\m\n}^{(m)} &:=& - \, \fr 2 {\sq{-\hg}}\, \fr \d {\d\hg^{\m\n}}
\Big[\sq{-\hg}\, \hLm\Big] \,\,, 
\label{A-EMT_m} \\
\hT_{\m\n}^{(\vph)} &:=& \fr 2 {\sq{-\hg}}\, \fr \d {\d\hg^{\m\n}}
\le[\sq{-\hg} \le\{\fr 1 2\, \hg^{\m\n} \pa_\m \vph \, \pa_\n \vph +\, 
U(\vph)\ri\}\ri] \nn\\
&=& \pa_\m \vph \, \pa_\n \vph \,-\, \fr 1 2\, \hg_{\m\n} 
\Big[\hg^{\a\b} \pa_\a \vph \pa_\b \vph \,+\, 2 U(\vph)\Big] \,, 
\label{A-EMT_f}
\eea
the variation of the action (\ref{A-E-ac}) leads to the relation
\be \label{A-consv}
\hnab^\m \,\hT^{(m)}_{\m\n} =\, -\, \hnab^\m \, \hT^{(\vph)}_{\m\n} 
=\, \cQ_\n \,, 
\ee
where 
\be \label{int-vec}
\cQ_\n =\, -\, \k \, n \, \hT^{(m)} \, \pa_\n \vph \,\,,
\ee
and $\,\hnab^\m = \hg^{\m\n} \hnab_\n \,$ and $\, \hT^{(m)} = \hg^{\m\n} 
\hT^{(m)}_{\m\n} \,$ are, respectively, the covariant derivative and the 
trace of the matter energy-momentum tensor in the Einstein frame. 

Eq.\,(\ref{A-consv}) shows that $\hT_{\m\n}^{(m)}$ and $\hT_{\m\n}^{(\vph)}$
are not conserved, i.e. there exists an interaction between matter and the 
field $\vph$, the extent of which is determined by the vector $\cQ_\n$. This
is nothing but the consequence of the $\vph$-dependence of the matter 
Lagrangian density $\hLm$, defined by Eq.\,(\ref{A-E-Lm}), in the Einstein 
frame. Hence, $\hLm$ is strictly speaking, an interaction Lagrangian density. 
In the standard cosmological framework, such an interaction can inevitably 
be looked upon as the DE-matter (DEM) interaction, under the legitimate 
assumption that the field $\vph$ sources the DE component of the universe.
%


\end{document}